**STXM-XANES and TEM analysis of UltraCarbonaceous Antarctic MicroMeteorites (UCAMMs)**

B. Guérin[1], C. Engrand[1*], C. Le Guillou[2], H. Leroux[2], E. Dartois[3], J. Duprat[4], S. Bernard[4], K. Benzerara[4], L. Delauche[1], D. Troadec[5].

[1]Université Paris-Saclay, CNRS, IJCLab, 91405, Orsay, France (cecile.engrand@ijclab.in2p3.fr), [2]Univ. Lille, CNRS, INRAE, Centrale Lille, UMR 8207 - UMET, F-59000 Lille, [3]ISMO, CNRS/Univ. Paris Saclay, 91405 Orsay Campus, France, [4]MPMC, CNRS/Sorbonne Université/MNHN, Case postale 115, 4 place Jussieu, 75252 Paris Cedex 05, France, [5]IEMN, CNRS/Univ. Lille, 5655 Villeneuve d'Ascq, France.

*corresponding author

**Abstract**: The Concordia micrometeorite collection contains rare (~1%) particles, which are unusually dominated by organic matter. These UltraCarbonaceous Antarctic MicroMeteorites (UCAMMs) are of probable cometary origin, and are thought to form in the outer regions of the protoplanetary disk. This work presents a multi-technique characterization of FIB (Focus-Ion Beam) sections of eight UCAMMs. Their organic and mineral contents have been investigated using Scanning Transmission X-ray Microscopy coupled with X-ray Absorption Near Edge Structure (STXM) and (Scanning)Transmission Electronic Microscopy ((S)TEM). The analyses revealed the presence of 3 different phases of organic matter in UCAMMs and small mineral aggregates distributed in an extended organic matrix. Organic characterization revealed that two of the organic phases in UCAMMs share clear spectral features with those of insoluble organic matter (IOM) of carbonaceous chondrites and of pristine organic from cometary grains returned by the STARDUST mission. The third organic phase is very N-rich (N/C atomic ratios up to 0.22) and has only a few equivalent in extraterrestrial organic matter analyzed so far, such as in cometary particles collected during the STARDUST mission (De Gregorio et al., 2011)}, Interplanetary Dust Particles (IDPs) (Aleon et al., 2003) and UCAMMs collected near the Dome Fuji Station (Yabuta et al., 2017) . This N-rich organic matter was probably formed by irradiation of N-rich ices in the outer regions of the protoplanetary disk. UCAMMs contain variable amounts of minerals, in this work only 4 FIB sections contained minerals that were characterized by (S)TEM. The minerals embedded in the organic matter of UCAMMs are constituted of crystalline Mg-rich silicates, Fe(Ni) sulfides and Fe oxides, which can be cemented in a Si-rich groundmass. Hypocrystalline mineral assemblages are also observed. Glassy phases showing morphological resemblance to GEMS are also found in some UCAMMs. This mineralogy is mainly consistent with that of previous UCAMM analyses, and is close to what is observed in Chondritic Porous Interplanetary Dust Particles (CP-IDPs), which are also considered as cometary dust particles. A mineral exhibiting a phyllosilicate texture has been observed in this work, and raises the question of the possibility of aqueous alteration on comets. Overall, UCAMMs' mineralogical and organic characterization implies a complex formation history of UCAMMs, and the presence of large scale radial mixing in the early solar system, in order to transport their mineral components to the outer parts of the protoplanetary disk.

# 1 Introduction

The flux of extraterrestrial material and organic compounds on Earth nowadays is mostly carried by the smallest objects falling on Earth, micrometeorites (Duprat et al., 2006; Engrand et al., 2017; Love and Brownlee, 1993; Rojas et al., 2021; Taylor et al., 1998). These submillimetric dust particles originate from various locations, probing regions from the inner and outer solar system. Micrometeorites represent the largest part of this flux of extraterrestrial material on Earth. Most of this dust is thought to derive from Jupiter-family comets, Kuiper belt and Oort Cloud comets, with a minor contribution (~20%) from asteroids (Nesvorný et al., 2010; Nesvorný et al., 2011; Poppe et al., 2011; Poppe, 2016; Rojas et al., 2021).

Micrometeorites presented in this work have been collected in Antarctica near the French-Italian Concordia station located at Dome C. The recovery of extraterrestrial material on the Antarctic continent allows the collection of well-preserved samples. Dome C vicinity is located inland, and micrometeorites are collected from 4-meter deep trenches to avoid human surface contamination. They are extracted from snow accumulated before human activity was present at Dome C due to the construction of the Concordia station. This context allows a unique preservation of extraterrestrial particles as temperature is low (-45°C in the trench) and mechanical stress is relatively weak as they are embedded in snow. The extraction method of micrometeorites has been described in Duprat et al. (2007). Among extracted extraterrestrial material, rare carbon-rich micrometeorites have been identified in several micrometeorites collections (Dobrică et al., 2012; Duprat et al., 2010; Nakamura et al., 2005; Yabuta et al., 2017) and are referred to as UltraCarbonaceous Antarctic MicroMeteorites (UCAMMs).

UCAMMs are dominated by organic matter, with a minor mineral contribution. Organic phases in UCAMMs can measure up to several tens of µm in size and represent more than 90% in volume of the micrometeorites. Previous analyses of UCAMMs indicate the organic content is N-rich and has an elevated D/H ratio (Dartois et al., 2018; Duprat et al., 2010). The organic matter (OM) of UCAMMs can be accessible without having to concentrate the organic matter, while on occasions mineral matrix meteorites of would need to be dissolved by acid treatments to concentrate the IOM for its detection It also allows keeping the 3D structure of the particles intact and better preserve and investigate the relationship between their organic and mineral phases. The large and accessible proportions of organic material present in UCAMMs can give new insight into organic matter structure at the nanometer scale as well as potential hints for its formation process.

In this work, we report on the organic and mineral analyses of fragments of 8 different UCAMMs by scanning transmission X-ray microspectroscopy (STXM-XANES) and focus on the mineralogy of 3 UCAMMs, as studied by transmission electron microscopy (TEM). We discuss potential implications for the formation and evolution of UCAMMs.

# 2 Material and Methods

## 2.1 Sample preparation and focused ion beam (FIB) sections

Fragments from eight UCAMMs collected in Antarctica snow, near the Concordia station (Duprat et al., 2007) were deposited on an aluminum stub covered with carbon conductive tape and carbon-coated. They were subsequently analyzed by a LEO 1530 scanning electron microscope (SEM) equipped with an energy dispersive X-ray spectrometer (EDX). Backscattered electron images of the UCAMMs fragments are presented in Figure 1 and the list of the analyzed fragments and the initial sizes of the corresponding particles before fragmentation is given in Table 1. Eight sections with a 100 to 150 nm thickness were prepared by Focused Ion Beam (FIB) at the Renatech facility at the Institut d'Electronique, de Microélectronique et de Nanotechnologie (IEMN, Lille), directly from the fragment deposited on the carbon tape. The FIB technique allows the extraction of ultrathin sections from the UCAMMs fragments without alteration of the textural integrity of the sample. Compared to ultramicrotome preparations, the FIB sections prevent nanoscale deformations sometimes encountered in microsectioned materials. The FIB sections were extracted with gallium ion milling. Although the milling was performed at low current and voltage in order to minimize the potential induced artifacts, slight implantation of Ga nanoparticles and sputtered material on the surface of the extracted section are sometimes visible. Nevertheless, the controlled low current prevented local compositional change or extensive formation of amorphous layers.

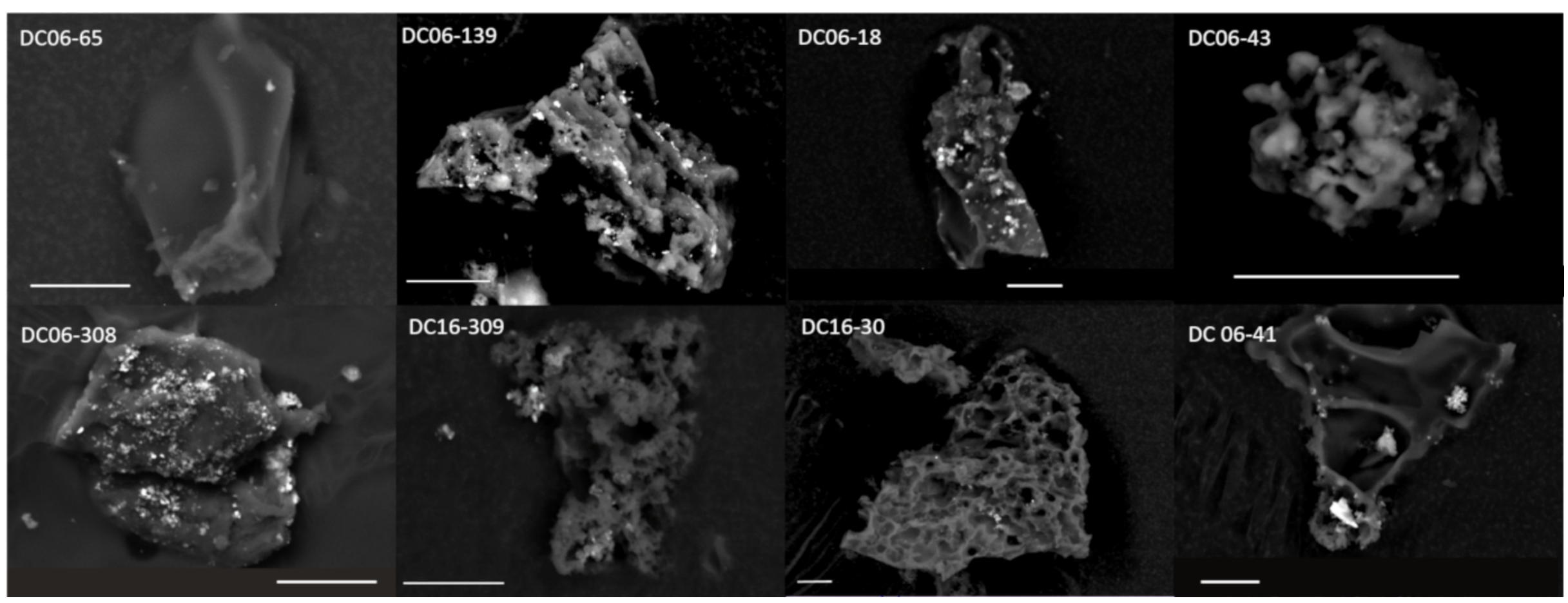

*Figure 1: Backscattered electron images of the fragments 8 UCAMMs mentioned in this paper. Scale bar is 10 µm.*

*Table 1. List of the UCAMMs analyzed, with corresponding sizes of the original micrometeorite. FIB sections were extracted from the fragment present on the stub, as shown in Figure 1.*

| UCAMMs | Full name | Initial size (µm) | Fragment stub number |
|---|---|---|---|
| DC06-308 | DC06_11_308 | 65x91 | ST11-13-07 |
| DC06-41 | DC06_07_41 | 80x200 | ST09_02_26 |
| DC16-30 | DC16_11_30 | 150x150 | ST16-09-30 |
| DC16-309 | DC16_14_309 | 60x70 | ST17-06-26 |
| DC06-139 | DC06_04_139 | 50x90 | ST14-03-09 |
| DC06-43 | DC06_04_43 | 24x28 | ST08_36_19 |
| DC06-18 | DC06_07_18 | 87x53 | ST09_02_01 |
| DC06-65 | DC06_05_65 | 44x33 | ST08_31_17 |

## 2.2 Scanning transmission X-ray microscopy-based X-ray Absorption Near-Edge Structure (STXM-XANES) spectroscopy

STXM-XANES is a transmission technique based on a monochromatic X-ray beam, generated by synchrotron radiation. It allows both microscopic imaging and spectroscopic measurements at the scale of a few tens of nanometers. The spectroscopy corresponds to X-ray absorption near-edge structure (XANES), and provides access to e.g. the chemical bonding configuration of the atoms in the sample. Depending on the speciation at the K-edge of either C-, N- and O- atoms in the different phases of the sample section, differential absorption will occur for XANES below the ionization energy of the different elements. In this study, we report data on 8 UCAMMs sections analyzed by STXM-XANES at the C-, and N K-edges. STXM-XANES data were collected on the beamline 11.0.2 at the Advanced Light Source (ALS) in Berkeley (for DC06-18 and DC06-65) and on the HERMES beam line at the synchrotron SOLEIL (for the other samples). At SOLEIL, a low-pressure flow of $O_2$ is added to the beamline that allows removal by photo-oxidation of carbon contamination on the optics. Energy calibrations are performed before the measurements, using as references the 3p Rydberg state peak at 296.96 eV for the C K-edge of gaseous $CO_2$ and the 1s -> π* photo-absorption resonance of gaseous $N_2$ at 400.8 eV for the N K-edge.

Stacks of absorption images were collected at many different energies across the carbon K-edge (270 eV - 355 eV). A 100 meV resolution was achieved between 282 and 295 eV, i.e. covering the pre-edge and main absorption fine structure. A 1 eV resolution was used between 282 eV and 295 eV, i.e. the pre- and post-edge energy ranges. For nitrogen, image stacks were collected over the energy range from 370 eV to 449 eV with a 100 meV resolution for the fine structure energy range and 1 eV for the pre and post edges. The pixel size measured between 30 and 50 nm depending on the chosen total size of the map and the dwell time was set to 1 to 2ms per pixel and per energy steps. The energies of maximum absorption are related to the electronic environment of the atoms which itself depends on their chemical bonding, and is used to identify the different functional groups. By measuring C and N simultaneously, we get access to the N/C ratio of the samples. The recorded data are processed with the aXis2000 software (ver. Oct 2018) to extract spectra and images (http://unicorn.mcmaster.ca/aXis2000.html). First, each UCAMMs FIB section was mapped at selected energies corresponding to the background absorption and major functional groups (280 eV, 285 eV, 288.6 eV, 290.3 eV), allowing a global identification of regions of interest in the sections. This method allows fast acquisitions, large scale, and high spatial resolution (pixel size = 50nm). Detailed sample stacks were then acquired with high-energy resolution (0.1 eV), and better spatial resolution (35 nm) over smaller area(s) of the section. Slight shifts in energy evidenced by the C (+0.4 eV) and N (-0.1 eV) calibration have also been corrected for, using the "Calibrate" function of aXis2000.

After the hyperspectral data is acquired, the "stack fit" procedure of the aXis software is used to deconvolve the data and to identify the areas with similar chemical signatures. One or several spectra are given as input and a linear combination is performed, so that all the pixels sharing similar spectral signature are identified as one component. The procedure is repeated until the whole hyperspectral data can be represented by a linear combination of the identified components.

The XANES spectra are eventually processed using the Quantorxs code (Le Guillou et al., 2018). Data processing with Quantorxs enables estimating the N/C atomic ratios and also provides a relative quantification of some functional groups abundances. The N/C quantification is carried out by fitting a power law to the pre-edge region and then by integrating the spectra from the pre-edge energy (282 eV for C, 397 eV for N) to the mean ionization energy (291.5 eV for C, 406.5 eV for N). The resulting integrated spectra were weighted by the absorption factors of carbon (3.67) and nitrogen (3.81) respectively. These factors correspond to the imaginary part (f2) of the complex atomic scattering factor at the ionization energy of the atoms (291.5 eV for C and 406.5 eV for N) as described by Henke et al. (1982; 1993). Once corrected by such absorption factors, a relative quantification of the N/C atomic ratio for each spectrum is obtained with an estimated error of $1\sigma_{N/C}$= 0.02.

## 2.3 Transmission Electronic Microscopy (TEM) analyses

The mineral assemblages of the same FIB sections were studied using transmission electron microscopy (TEM). The investigations have been performed on a ThermoFisher Scientific Tecnai G2 20 (DC06-41, DC06-43, DC06-18, DC06-65) and ThermoFisher Scientific Titan Themis 300 (DC06-41, DC06-43, DC16-309, DC06-308, DC16-139) at the electron microscopy facility of the Université de Lille. The data on the Titan Themis were acquired in the scanning TEM mode (STEM), operating at 300keV. These analyses allowed the study of mineral assemblages in UCAMMs at spatial resolution down to the nanometer scale. The Tecnai microscope is equipped with a Bruker energy dispersive X-ray spectrometer (EDX). The Titan Themis microscope is equipped with a four-quadrant, windowless, super-X SDD EDX detector). Electron diffraction has also been performed on complex mineral assemblages of DC06-308. EDX data acquired with the Titan Themis microscope were processed thanks to the hyperspy software (doi : 10.5281/zenodo.4294676) and those acquired on the Tecnai microscope were processed with the Bruker Esprit® software. For both we used experimentally determined k-factors for major elements (O, Mg, Al, Si, Fe (K and L lines), S, Ca, K). For EDX data acquired with the Titan Themis microscope, hyperspectral EDX maps were acquired, which contain an EDX spectrum (from 0.15 to 20 keV) for each pixel. In order to optimize the quantification, the different hyperspectral maps have been rebinned to increase the signal/noise ratio of the EDX spectra. To reveal the different phases, we applied singular value decomposition (SVD) and extracted the end member spectra from the corresponding regions of interest. For the hyperspectral EDX maps acquired on the Tecnai microscope, the areas of interest were selected manually. To quantify the EDX spectrum of each identified phases and perform accurate absorption correction, the "local" thickness and density of the sample must be constrained. This has been done thanks to the constraints provided by the presence of the Fe-K and Fe-L X-ray lines of Fe-rich phases (Le Guillou et al., 2018; Zanetta et al., 2019). As absorption differs for the two Fe X-ray lines, and because the quantification using either the K or the L line should give the same result, the values of the thickness and density are iteratively explored until the equal quantification condition is met for each analyzed area in the FIB sections, STEM high-angle annular dark-field (HAADF) images were taken before hyperspectral EDX maps (Titan Themis) or TEM and STEM bright field (Tecnai).

# 3 Results

Here, we report on the characterization of the organic matter by STXM of 8 UCAMMs and new mineral analysis in FIB sections by STEM of UCAMMs DC06-308, DC16-309 and DC06-18. Figure 2 shows the FIB sections of the 8 UCAMMs analyzed, with the locations of the different regions analyzed by STXM-XANES and (S)TEM.

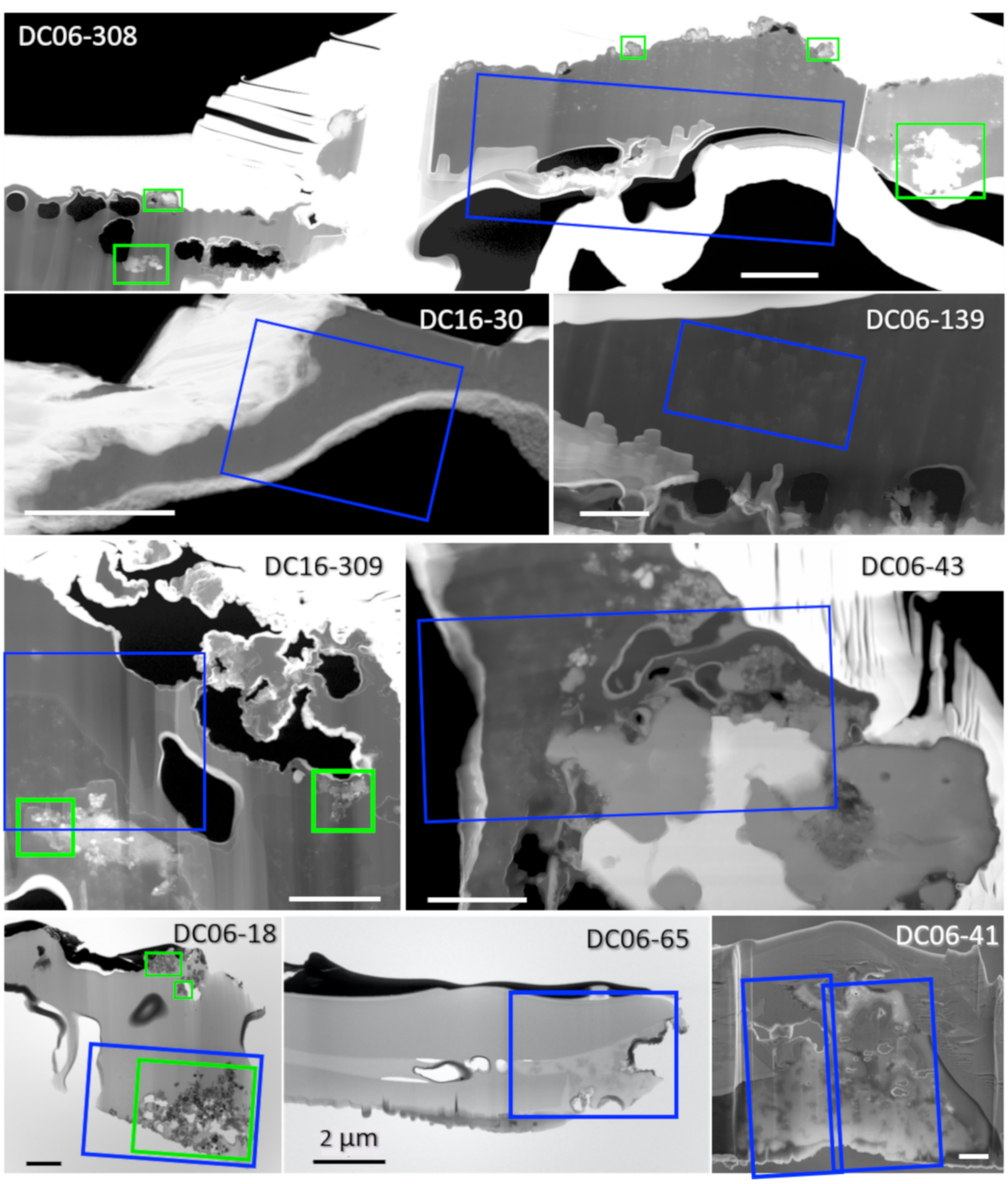

*Figure 2: Electron microscope images of the eight UCAMM FIB sections analyzed in this work: STEM High-angle annular dark field (HAADF) images for DC06-308, DC16-30, DC06-139, DC16-309, DC06-43); Bright field STEM images for DC06-18 and DC06-65, SEM image of DC06-41 at 5 kV (the FIB section was lost before STEM) . The platinum layer deposited on top of the sample during the FIB process appears in white in the HAADF images, in black in the bright field STEM images, and in smooth gray in the SEM image. The blue boxes show the locations of the STXM-XANES maps shown in Figure 3. The green boxes show the locations of mineralogy characterizations by STEM/EDX presented in this work. Scale bar is 2 µm.*

## 3.1 Characterization of the organic matter of UCAMMs

### 3.1.1 STXM-XANES analysis

The STXM-XANES hyperspectral maps were acquired at the C and N K-edges for the eight UCAMMs. Three main types of organic matter (OM) can be identified based on their different carbon and nitrogen speciation in the eight UCAMMs analyzed in this study. In Figure 3, the three identified types of OM are represented in blue, green and red in RGB images. Figure 4 and Figure 5 present the corresponding spectra at the carbon and nitrogen K-edges, respectively. The characteristic peaks of type I OM (in blue in Figure 3) are found at 284.8 eV (aromatic and olefinic groups, C=C), 286.4 eV (ketone and phenol, C=O) and 288.4 eV (carboxyl, O=C-O) (see Figure 4). Type II OM (in green in Figure 3) displays functional groups similar to type I OM although the first peak is shifted to 285 eV, indicating a larger contribution of aromatic with respect to olefinic carbon. The N/C atomic ratio calculated for type I and II OMs ranges between 0.01 and 0.05 ($1\sigma=0.02$) and is below detection limit in some cases. For type III OM, the main absorption peaks are located at 284.8 eV and 286.4 eV, respectively corresponding to aromatic carbon and nitrile (N≡C). Note that in type III OM spectra, there is no drop around 285.8 eV, corresponding to the contribution of imine (C=N), as seen in Figure 4 and Figure 5. The type III spectra also display a smaller peak at 284.4 eV (alkene, C=C) than for types I and II, and the spectra do not show dominant oxygen-related peaks, although a small C=O-related absorption could contribute to the observed continuum in the spectra. The carboxylic contribution is lower, as for type II OM. The atomic N/C atomic ratios for type III OM are higher than for type I and II OMs ($0.07 \pm 0.02 < N/C_{III} < 0.20 \pm 0.02$, with an outlier at $N/C_{III} = 0.03 \pm 0.02$ for DC16-30 – see 4.3). Discrimination between nitrile and ketone absorption, both located at around 286.4 eV in the carbon K-edge spectrum, is based on the STXM-XANES spectrum at the nitrogen K-edge (370 eV - 449 eV). If a significant nitrile absorption peak is visible at the nitrogen edge, the 286.4 eV peak is attributed to nitrile although a fraction of the absorption could be related to a minor C=O contribution. Type I and II OM in UCAMMs do not show any nitrogen absorption, pointing toward N-content of this phase below the detection limit. For type III OM spectra, the main peaks are located at 398.8 eV and 399.8 eV and correspond to imine and nitrile absorption (Figure 5). Another smaller peak can be observed at 401.5 eV and can be attributed to amide group. The N/C atomic ratios of the different organic matter types and the presence (or not) of mineral phases in the FIB section of each UCAMMs are summarized in Table 2.

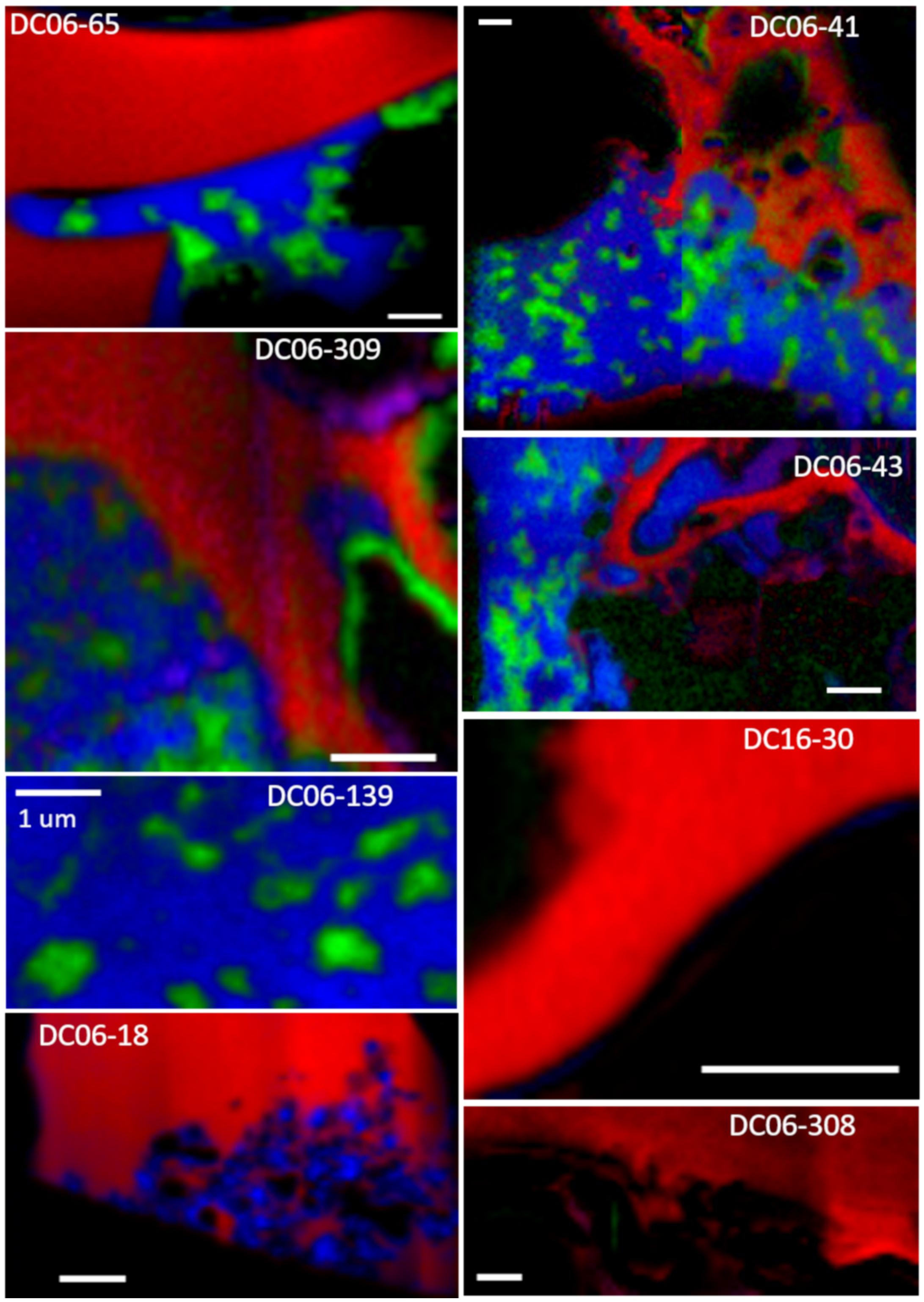

*Figure 3: RGB images obtained by deconvolving the hyperspectral STXM-XANES maps in the eight UCAMMs. Three organic phases are found: type I (in blue), type II (in green), and type III (in red). Scale bar is 1µm.*

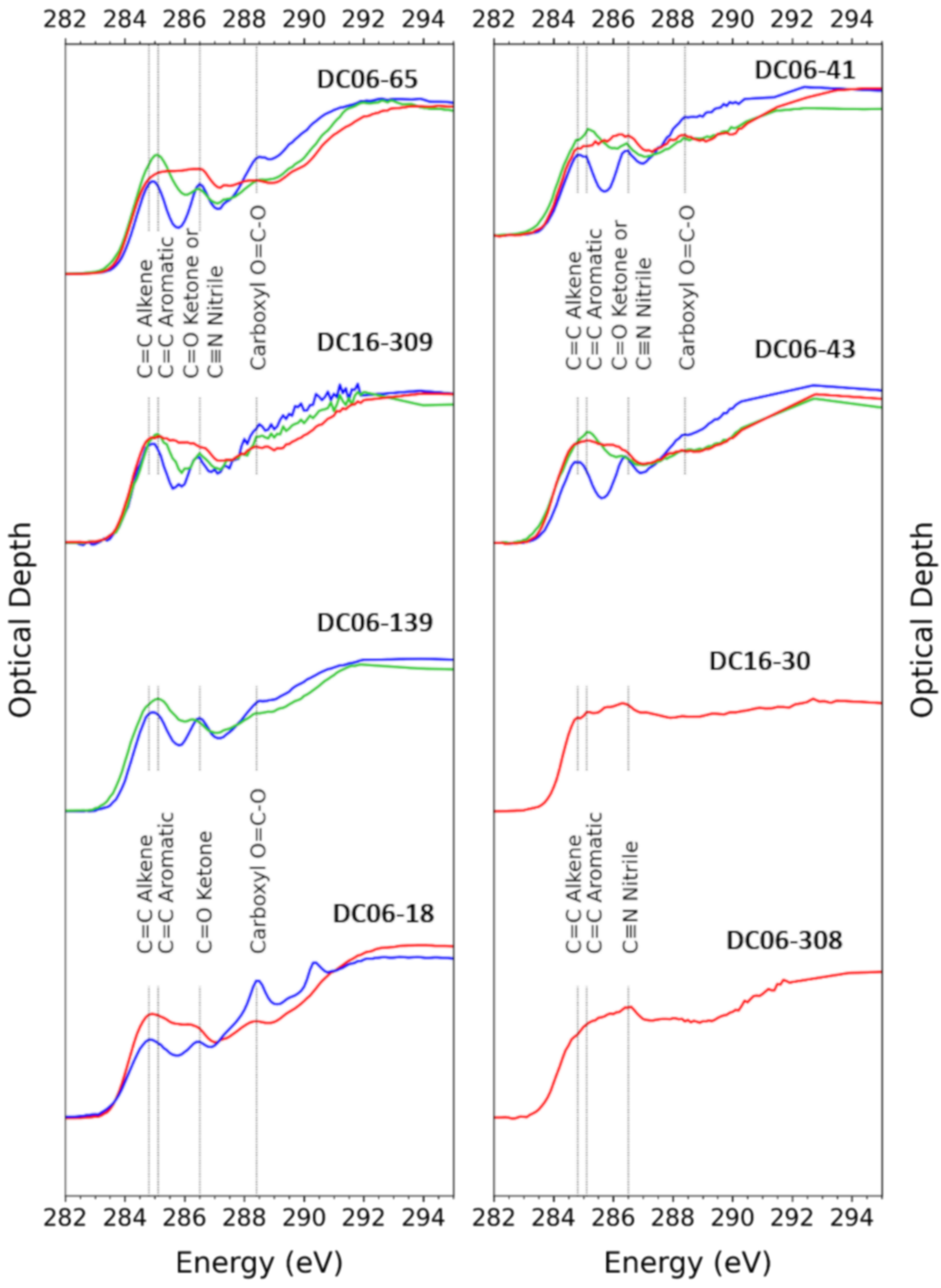

*Figure 4: Carbon K-edge XANES spectra for the three different types of OMs observed in the eight UCAMMs, as illustrated in Figure 3: type I (in blue), type II (in green), and type III (in red). The absorption peak attributions were made according to (De Gregorio et al., 2011).*

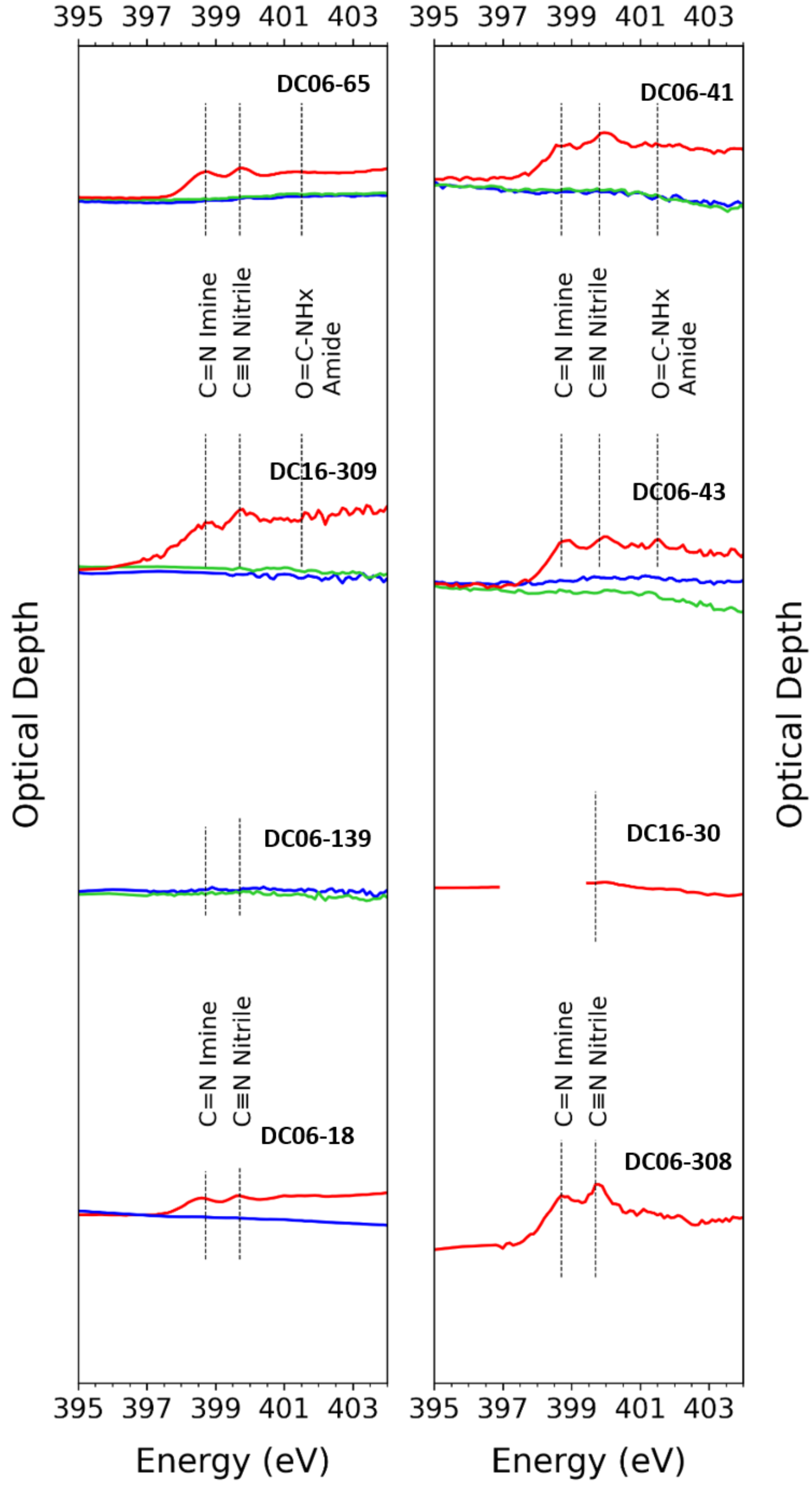

*Figure 5: Nitrogen K-edge XANES spectra for the three different types of OMs observed in the eight UCAMMs, as illustrated in Figure 3: type I (in blue), type II (in green), and type III (in red). The absorption peak attributions were made according to (De Gregorio et al., 2011).*

*Table 2: N/C atomic ratios of the different types of organic matter in the analyzed UCAMMs (from STXM-XANES quantification) and brief description of the mineral content observed (or not) within their FIB sections.*

*Table 2. N/C atomic ratios of the different types of organic matter in the analyzed UCAMMs (from STXM-XANES quantification) and brief description of the mineral content observed (or not) within their FIB sections.*

| | N/C atomic ratio (STXM-XANES) | | | Minerals in FIB sections (STEM analyses) |
|---|---|---|---|---|
| UCAMM | OM I | OM II | OM III | |
| DC06-308 | - | - | 0.20 | Isolated mineral patches |
| DC06-41 | 0.02 | 0.01 | 0.12 | Not analyzed. (FIB section damaged after STXM) |
| DC16-30 | - | - | 0.03 | No minerals |
| DC16-309 | 0.01 | $<0.01$ | 0.07 | Isolated mineral patch (Silicate, FeS, Phyllosilicate-like) |
| DC06-139 | 0.03 | 0.02 | - | No minerals |
| DC06-43 | 0.04 | 0.03 | 0.08 | Large mineral assemblage (Silicate, FeS), GEMS-like inclusions |
| DC06-18 | $<0.01$ | - | 0.11 | fine-grained mineral assemblages (Silicate, FeS), GEMS-like inclusions |
| DC06-65 | 0.03 | 0.05 | 0.20 | No minerals |

### 3.1.2 TEM imaging and EDX analysis of the organic matter in UCAMM DC16-309 and DC06-18

As the same FIB sections were used for both STXM-XANES and (S)TEM analysis, higher resolution images and EDX maps were acquired in the area analyzed by STXM-XANES, to characterize the fine structures and chemical heterogeneities at the nanometer scale. In Figure 6, we present a combination of RGB image with a detailed HAADF image of the three types of organic matter observed in the FIB section of UCAMM DC 16-309.

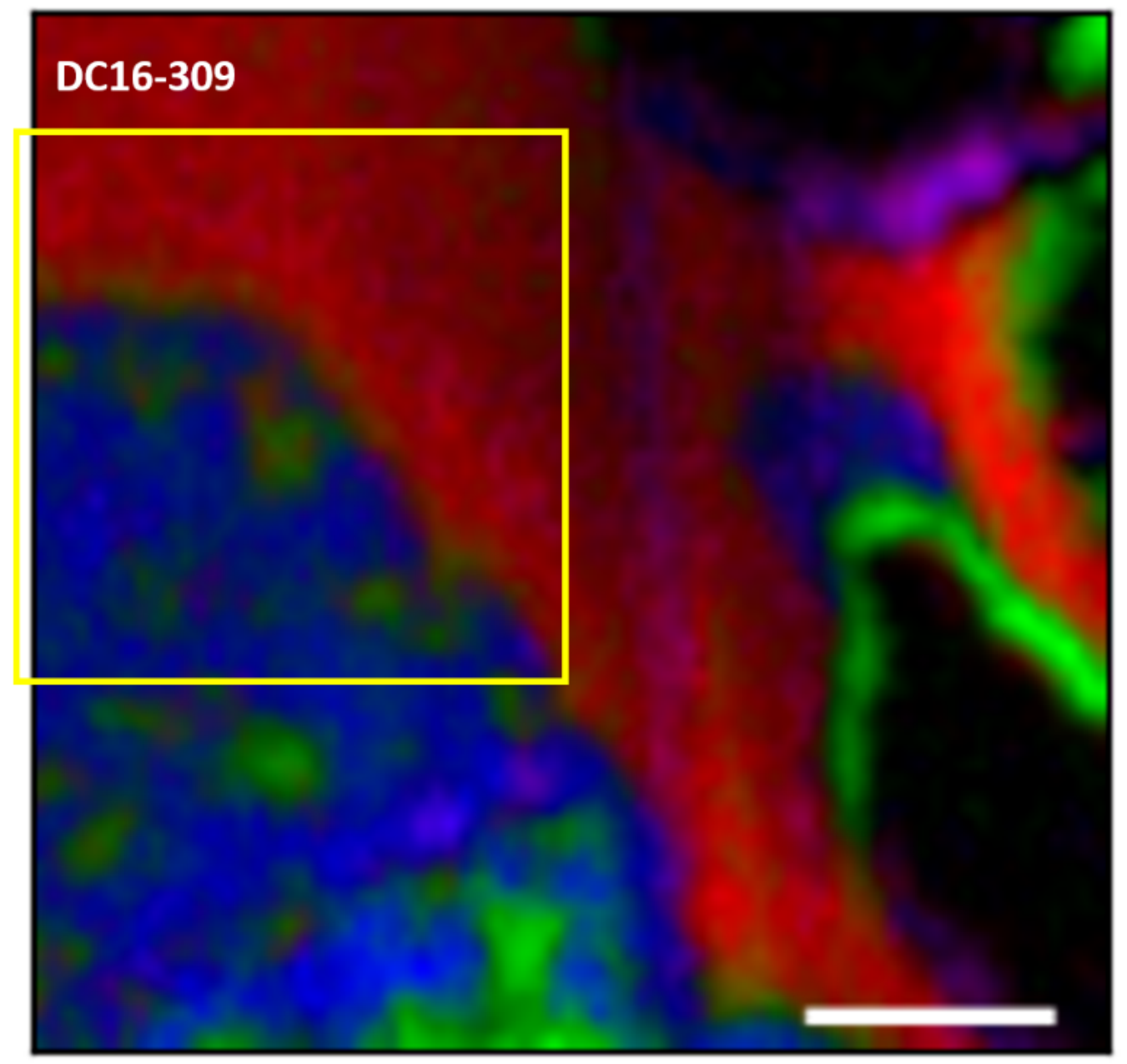

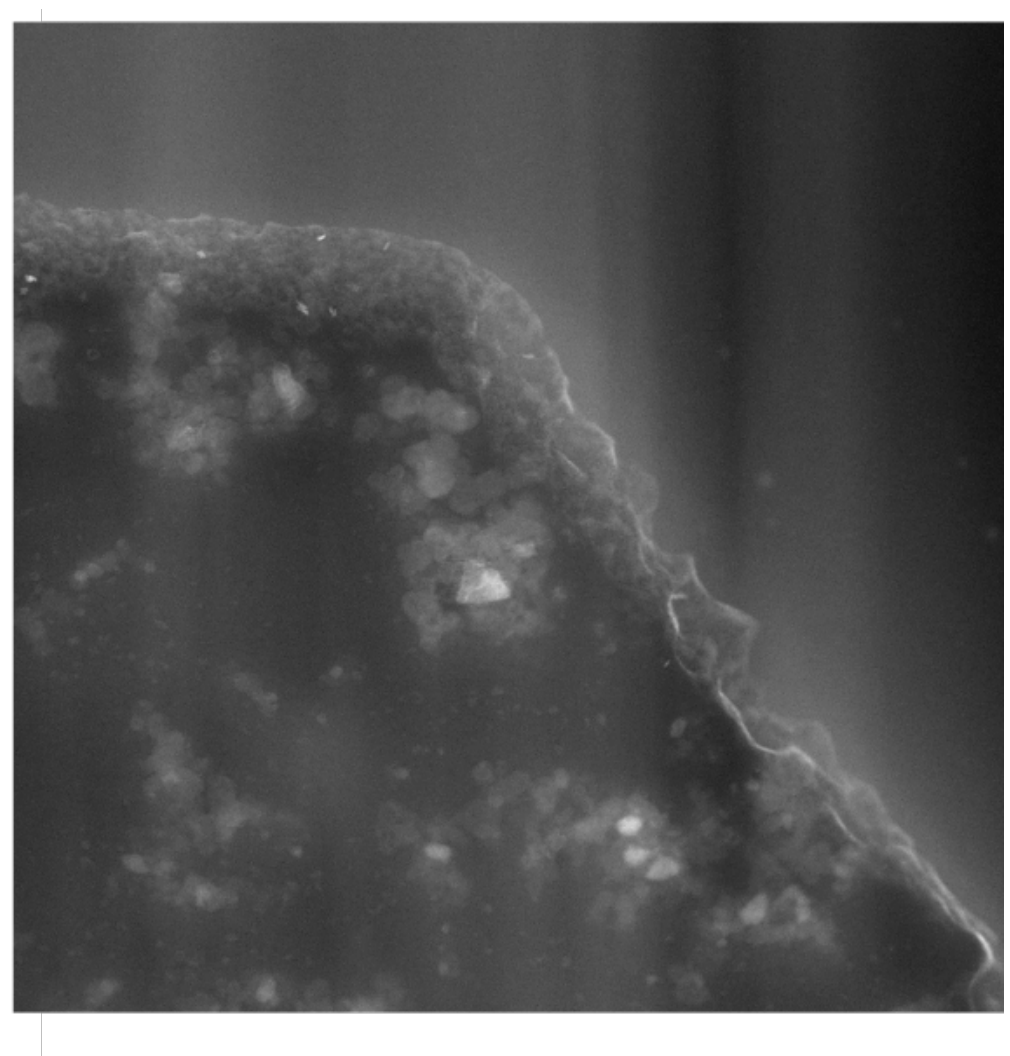

*Figure 6: (left) RGB color STXM-XANES image obtained by the spectral deconvolution of hyperspectral maps acquired at the C and N K-edges for UCAMM DC16-309 (scalebar is 1 µm). The three types of organic matter are present in the image. The smooth area of OM on the top of the image corresponds to type III OM (in red) while the bottom left structure corresponds to type I (in blue) containing the dusty patches of type II OM (in green). (Right) detail HAADF image of the top left area analyzed by STXM-XANES in UCAMM DC16-309 (yellow square).*

The three types of OM are not systematically found in all UCAMMs' FIB sections. STEM imaging of the organic matter in DC16-309 clearly shows that organic matter of types I, II and III observed in STXM-XANES have different textures (Figure 6). Type I and II OMs are generally associated together in the form of sub-micrometric patches ("dusty patches") of type II OM embedded in a type I OM (see DC06-41, DC16-309, DC06-139, DC06-43 and DC06-65 in Figure 3). In contrast, type III OM always presents a smooth texture. The STEM image in Figure 6 reveals the presence of a foamy-like layer between the organic matters of types I and III (detailed HAADF image). From the EDX map, it appears that the 'foamy-like' phase contains small amount of Fe (in red, Figure 7 middle). The presence of iron does not correlate with either oxygen or sulfur in this phase. The STXM-XANES analysis did not have the spatial resolution needed to extract the spectrum specific to the intermediate foamy layer. The elemental compositions inferred from the EDX spectrum of this layer indicates it is similar to type I OM. Areas resembling inorganic phases are also present in most of the dusty patches that represent type II OM. The EDX analysis of these inorganic phases indicates that they contain Na, S and a small amount of Si (Figure 7). EDX spectrum extracted from the region containing these inorganic phases do not display a clear stoichiometry but these brighter phases could correspond to small $Na_2S$ minerals.

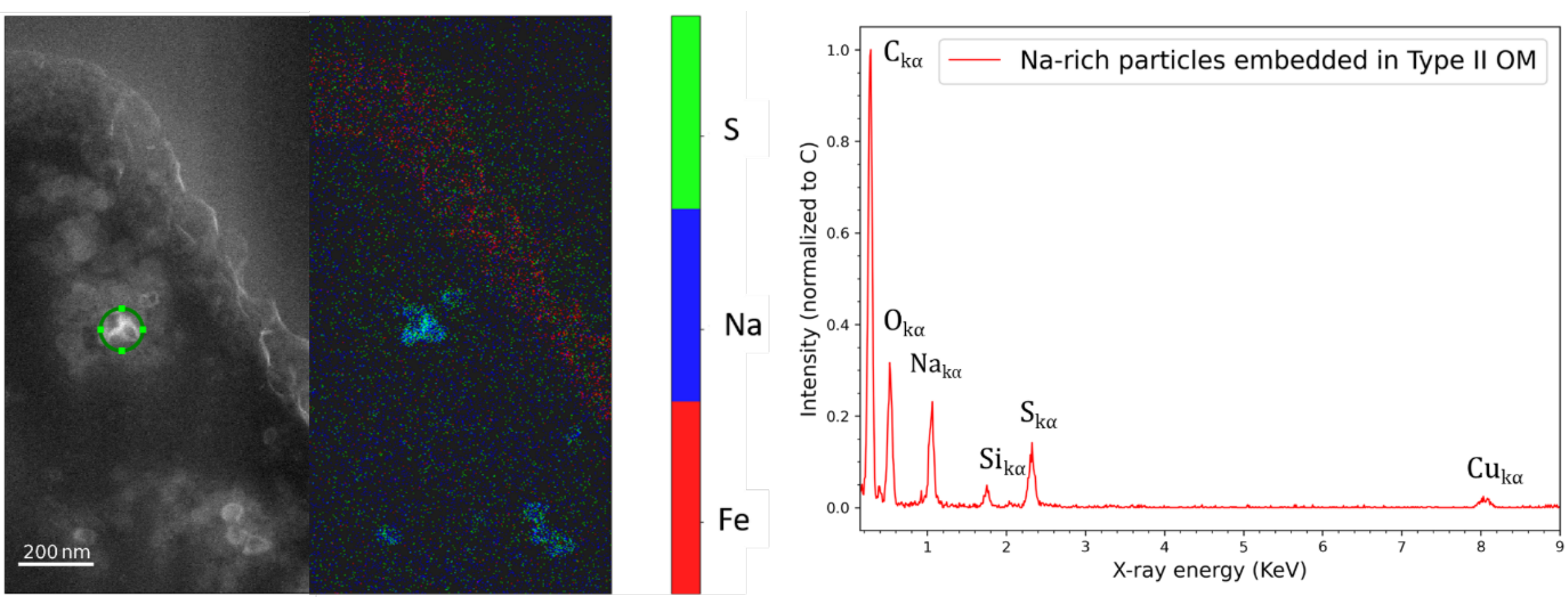

*Figure 7: (A) HAADF images of the EDX map of the organic matter of UCAMM DC16-309.The green circle corresponds to the location of the spectrum displayed in (c). (B) Elemental map showing Fe (in red), S (in green) and Na (in blue) based on X-ray lines intensities. Each element is normalized to its maximum for better visualization. (C) EDX spectrum extracted from the green ROI of the HAADF image normalized to carbon intensity. The Cu signal is due to the grid substrate.*

EDX spectra of the three types of OM have been acquired by selecting ROIs in the EDX hyperspectral maps, and are displayed in Figure 8, normalized to the maximum C intensity in each phase. The N-enrichment of the type III OM is visible in the spectrum in Figure 8, with a clear decrease of the N and O contents from type III to type I, and then to type II.

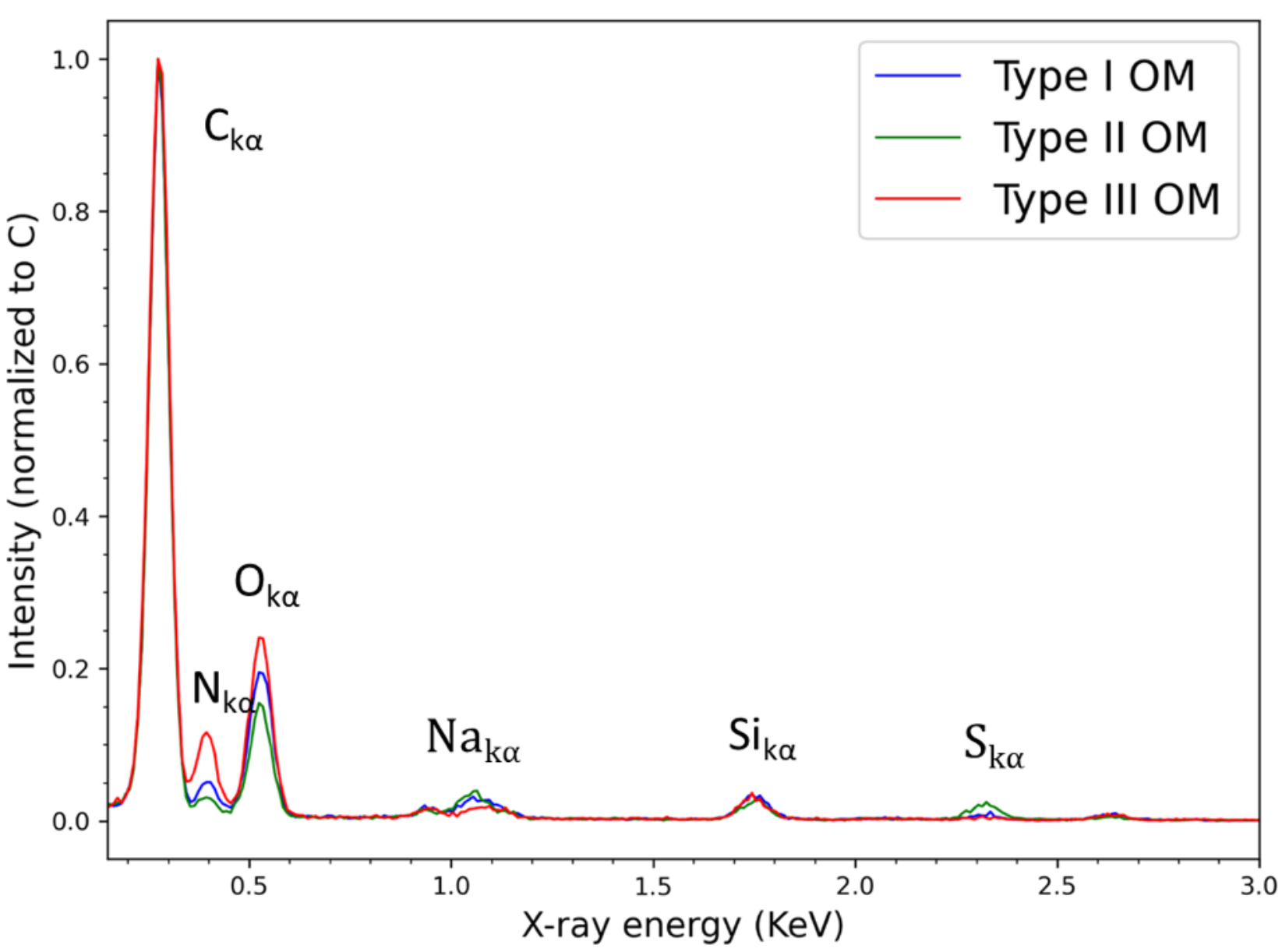

*Figure 8: EDX spectra of the three types of organic matter in DC16-309. Each spectrum has been normalized to its maximum carbon intensity for easier comparison.*

UCAMMs DC06-18 is mostly constituted of type III organic matter of that extends over most of the FIB section area. Although type III presents a smooth texture as well as rather uniform N/C elemental ratio over the FIB section, the chemical characteristics of the OM surrounding mineral assemblage shows unusual variation in the organic matter composition. The mineral assemblages are reminiscent of glass embedded metals and sulfides (GEMS) phases observed in IDPs. More specifically, gradients in composition around GEMS-like assemblages are observed in DC06-18 as illustrated in Figure 9 and Figure 10. Na and N increase toward the mineral assemblages at the 50 nm level. Organic matter appears to be found in small amounts inside the fine-grained inclusion shown in Figure 9, and exhibits a N/C atomic ratio that can reach a value of 0.5.

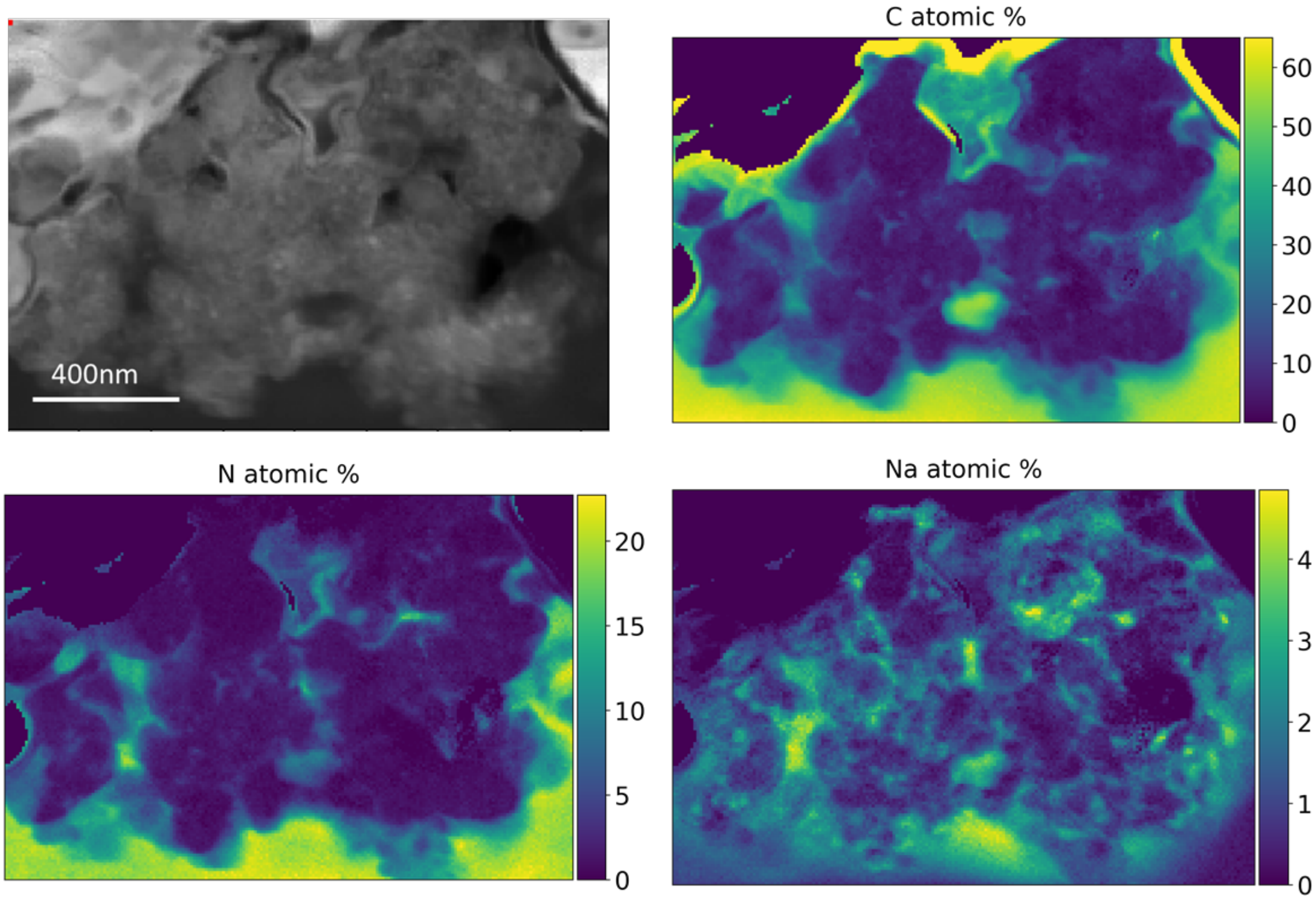

*Figure 9: HAADF image and quantified EDX map (at%) of a GEMS-like assemblage in UCAMM DC06-18, which is located at the top of the FIB section, probably close to the external surface of the UCAMM.*

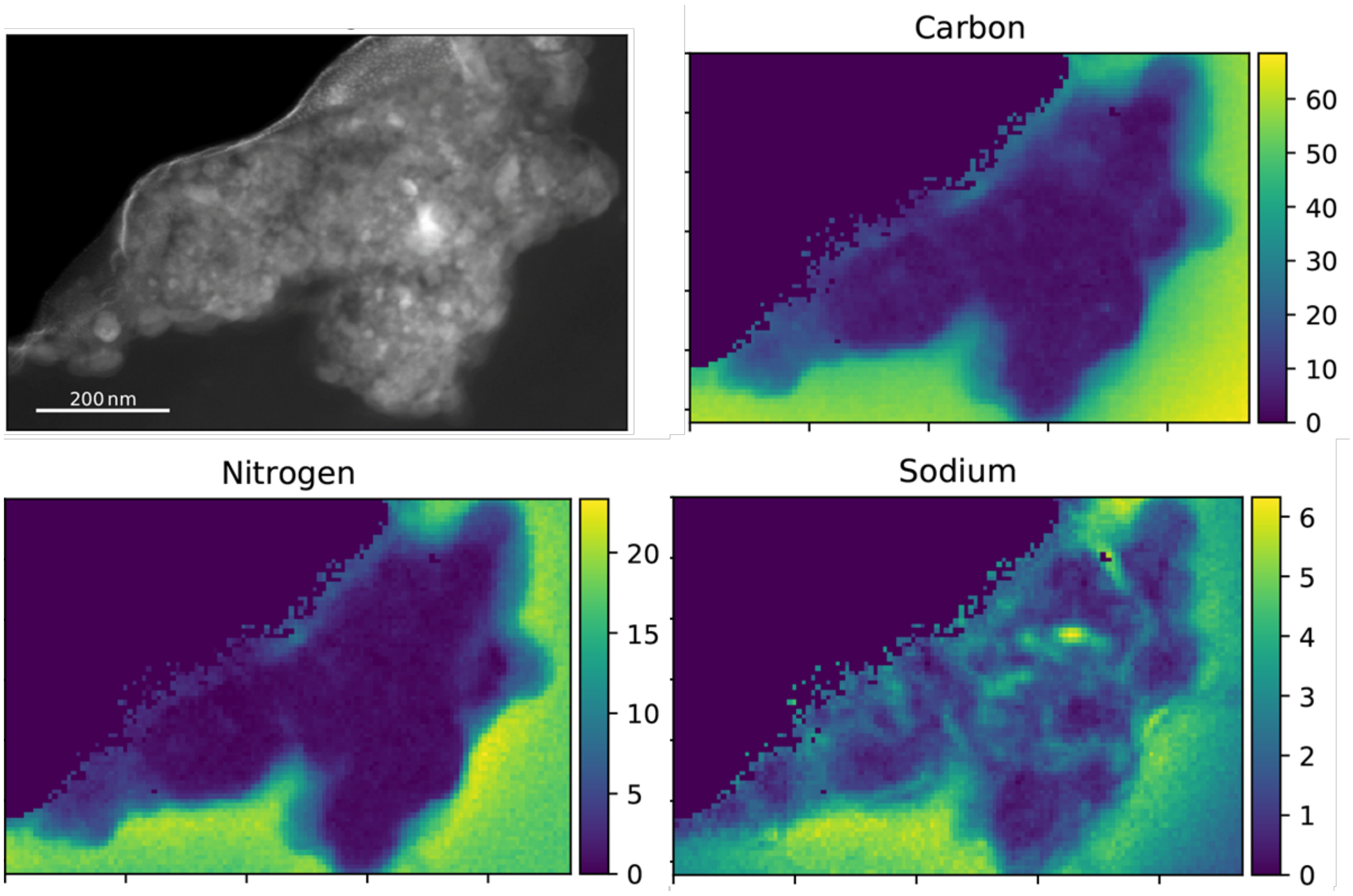

*Figure 10: HAADF image and quantified EDX map (at%) of a GEMS-like assemblage in UCAMM DC06-18, which is located at the top of the FIB section, adjacent to a vesicle.*

## 3.2 Inorganic phases analysis by (S)TEM of DC06-308, DC16-309 and DC06-18 UCAMM FIB sections

(S)TEM analysis revealed that only four out of the eight FIB sections contain inorganic inclusions embedded in an abundant carbonaceous matter (DC06-308, DC06-43, DC16-309 and DC06-18; Table 2). Overall, FIB sections are dominated by organic matter that extends over several microns. The FIB sections of DC16-30, DC06-139 and DC06-65 do not contain any mineral phases, and we could not investigate the mineralogy of DC06-41 by TEM, as the FIB section was lost after STXM-XANES analysis. Crystalline phases of UCAMMs reported by Dobrică et al. (2012), Engrand et al. (2015) and Charon et al. (2017), are mainly Mg-rich pyroxenes and olivines, Fe-sulfides and Fe-oxides. Crystalline grains are often decorated with Fe-sulfides and oxides, located at the edge of the primary minerals. Sulfides and oxides range from tens of nanometers to sub-micrometers in size while the assemblages of crystalline or hypo crystalline phases can reach several micrometers. We focus hereafter on the description of inorganic phases of DC06-308, DC16-309 and DC06-18.

### 3.2.1 Mineralogy of UCAMM DC06-308

The inorganic assemblages in DC06-308 FIB section are located in small and isolated patches. These assemblages are made of crystalline, hypocrystalline-like and amorphous components, sometimes associated together at a very fine scale (tens of nanometers, Figure 11). Crystalline and hypocrystalline-like phases are only found in type I OM while amorphous silicate can be present in all types of OM.

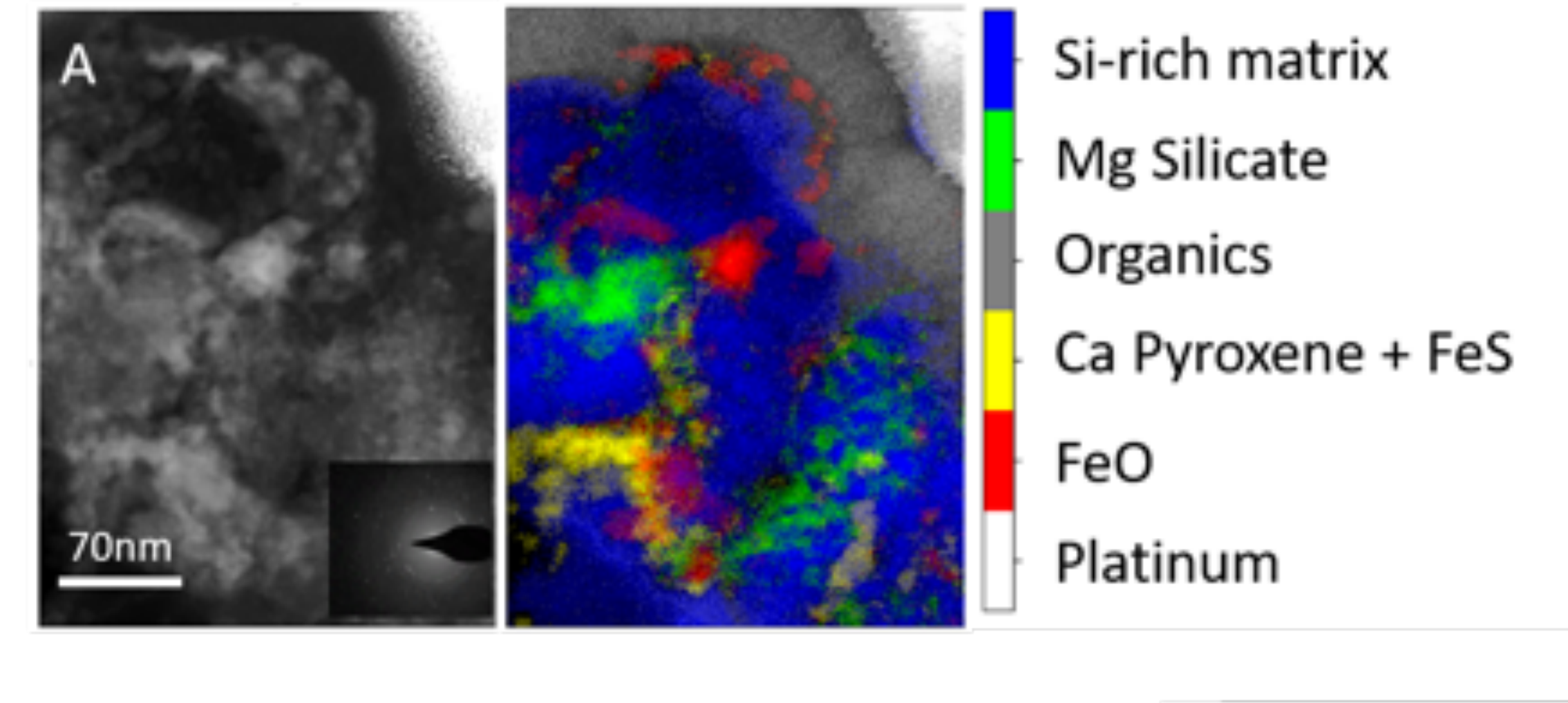

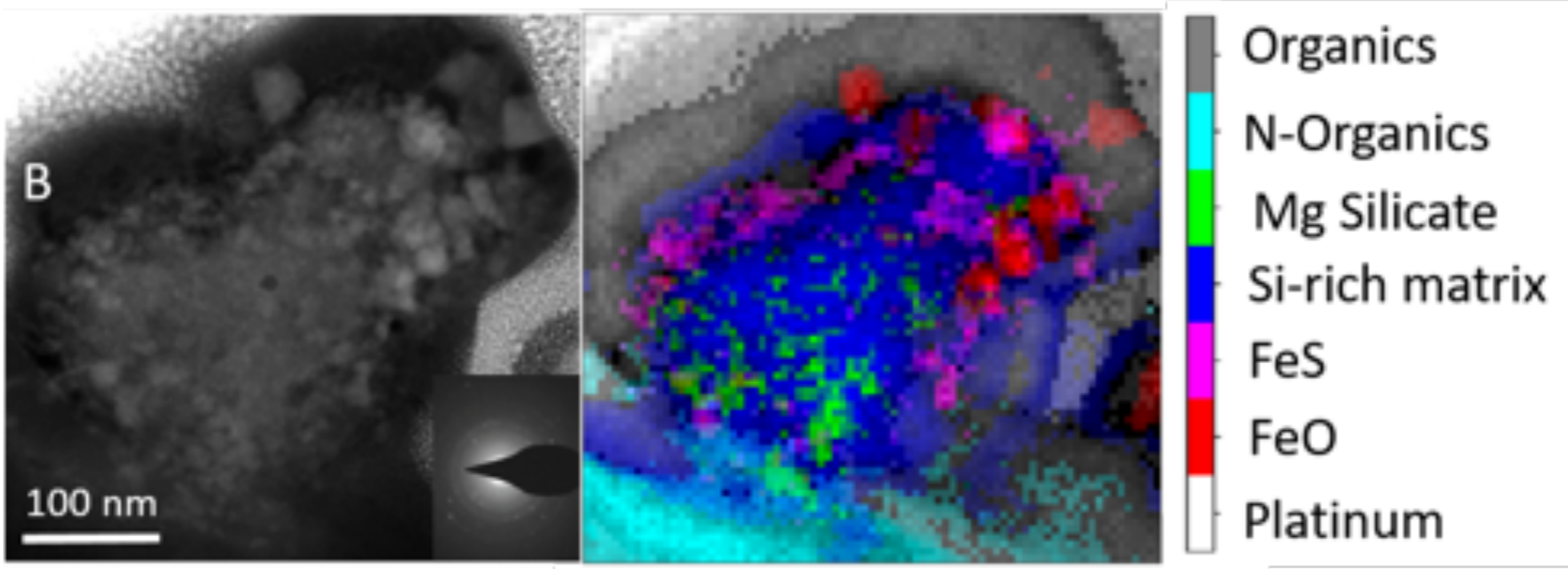

*Figure 11: (A and B, left) High Angle Annular Dark Field (HAADF) image of 2 mineral assemblages in DC06-308 (A and B, right) colored images corresponding to the EDX maps of the different phases identified by Principal Component Analysis (PCA). The phases are ordered by decreasing number of counts in the respective spectra*

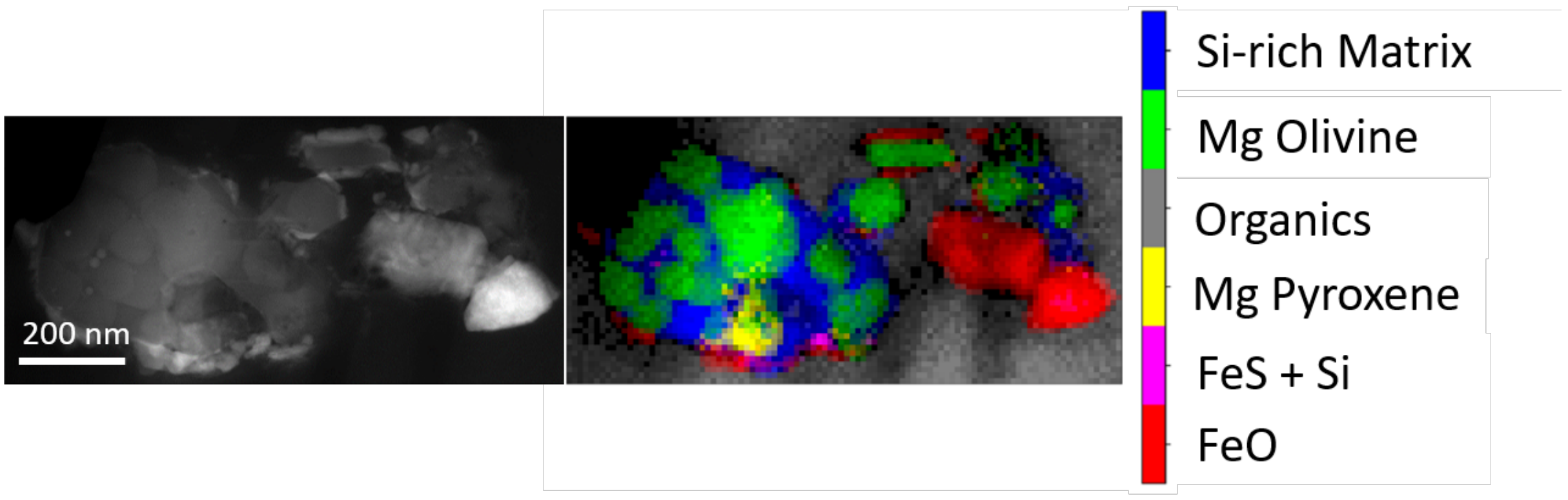

*Figure 12: HAADF image (right) and EDX phase distribution (left) of a crystalline assemblage in DC06-308 dominated by Mg-rich olivine grains cemented by an amorphous silica-rich matrix, with Fe oxides and Fe sulfides at the border of the object.*

DC06-308 revealed several mineral assemblages plotted in Figure 11 and Figure 12. The different phase maps produced by PCA indicate a fine grain texture, with several phases intimately mixed. All assemblages display a Si-rich groundmass (quoted here as “Si-rich matrix”) embedding Mg-rich olivines and Mg-rich low-Ca pyroxenes. A Ca-rich pyroxene has also been observed in one assemblage (Figure 11A), and seems closely associated with a Fe-sulfide. Fe oxides are usually decorating the edge of the assemblage for most mineral phases, along with Fe-sulfides in some cases (Figure 11B). Spatial heterogeneities can be present at a very small scale, down to a few nm. The mineral assemblage in Figure 12 shows a hypocrystalline-like texture consisting of Mg-olivines cemented by a Si-rich matrix.

Electron diffraction realized on two of the assemblages of DC06-308 shows complex pattern. One or two rings formed by diffraction of poorly crystalline material can be observed. The rings often contain diffractions spots indicating the presence of multiple small crystals with varied crystallographic orientations. However, unambiguous indexation of the patterns is made difficult by the composite nature of the assemblages. An indexation attempt based on the phases extracted from the hyperspectral deconvolution has been done and is summarized in table ?.

Compositions of the different phases from the mineral assemblages in DC06-308 are summarized in Figure 14. One silicate (in green in Figure 11A) is close to a forsterite composition with high Mg/(Fe+Mg) ($Fo_{95}$) ratio while another silicate phase (in green in Figure 11B) has a non-stoichiometric composition and a higher Fe concentration (Mg/(Fe+Mg) = 0.7). Silicates in Figure 12 are both Mg-rich and have compositions corresponding to forsterite and enstatite (in green and yellow, respectively). The matrix that cements the mineral grains is amorphous and shows Si-rich compositions (square symbols in Figure 14). Fe sulfides are Ni-poor and have a composition matching that of troilite (Figure 14B).

### 3.2.2 Mineralogy of UCAMM DC16-309

The analysis of DC16-309 revealed a mineral with a fibrous texture similar to that of a phyllosilicate with a total size of around 2.4 µm (Figure 13A). The global composition of this fibrous phase is close to that of the Si-rich matrix found in other assemblages except for a high Al/Si atomic ratio. A mineral with enstatite composition is also associated with the phyllosilicate-like phase. A mineral assemblage close to this fibrous phase contains large low-Ni Fe sulfides (hundreds of nanometers) and smaller Fe oxides (Figure 13A). This mineral assemblage is embedded in type I OM. Unfortunately, this FIB section was lost after the first STEM analysis and further characterization of the fibrous phase could not be realized.

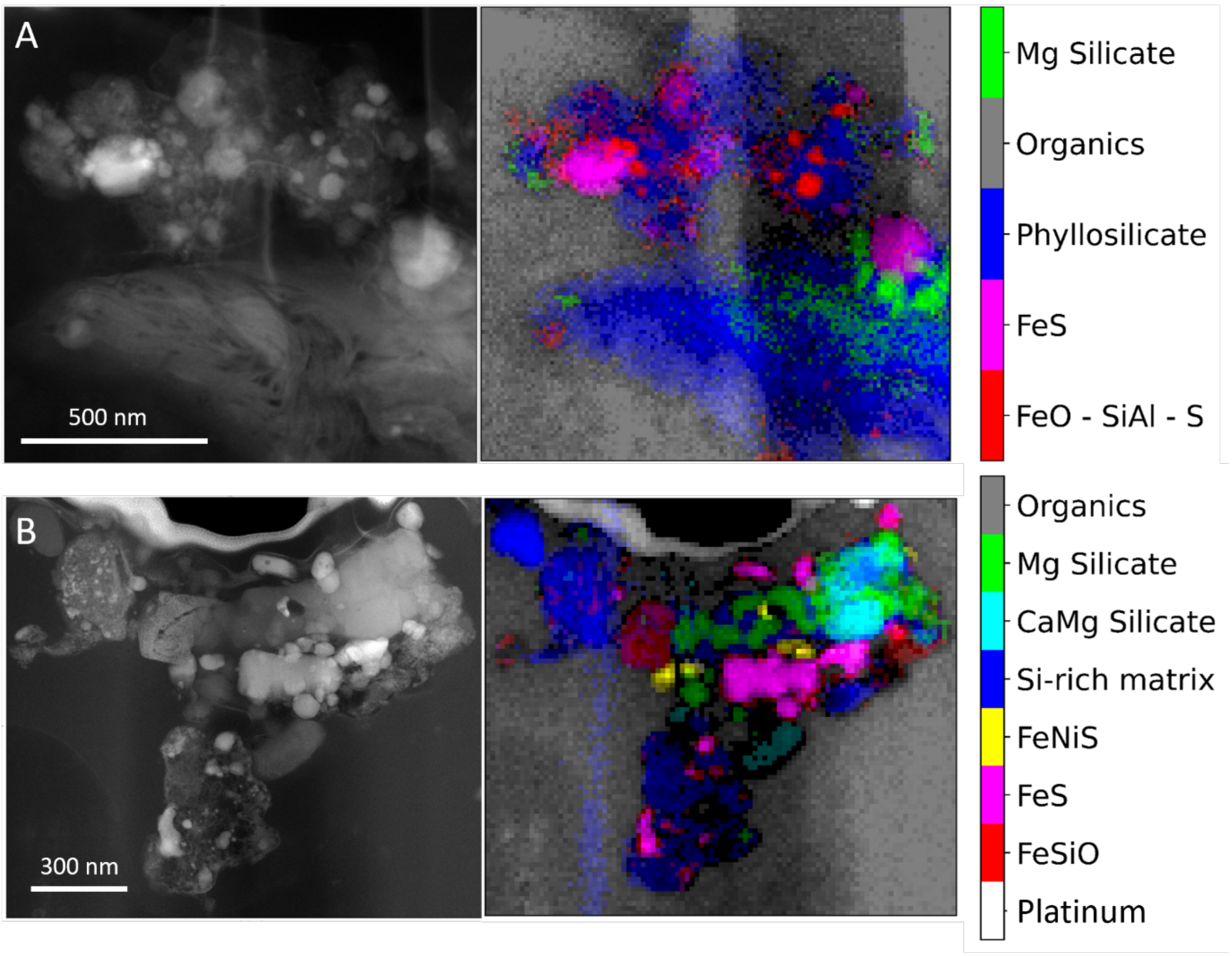

*Figure 13: (left) High angle annular dark field (HAADF) images and (right) colored images corresponding to the map of the different phases identified by Principal Component Analysis of the EDX hyperspectral maps. (A) Phyllosilicate-like phase (fibrous phase) decorated with a mineral assemblage. (B) Mineral assemblage located at the border of a vesicle. Both assemblages are embedded in the organic matter.*

DC16-309 exhibits another complex mineral assemblage (Figure 13B) made of Ca- and Mg-rich silicates, Mg-rich silicate, Fe-sulfides and oxides (~100 nm) and smaller FeNi sulfides (tens of nm). The Ca-rich and Mg-rich phases are closely associated. Fe oxides and sulfides are distributed at the edge of the assemblages or in the form of larger crystals.

The small sized Mg-rich silicates in DC16-309 display non-stoichiometric compositions possibly due to the sampling of several phases under the electron beam. The mineral associated with the phyllosilicate matches an enstatite composition (Figure 14). The nature of the Mg-rich silicate in the second assemblage (Figure 13B) is not clear (Figure 14). The matrix data (square symbols in Figure 14) show Si-rich compositions. Fe sulfides in DC16-309 present two distinct compositions. Fe sulfides associated with the phyllosilicate present low-Ni/Fe values with a composition close to that of pyrrhotite (Figure 14B). Similar compositions are found in Fe-sulfides of the second assemblages with low-Ni concentration. A second population of sulfides in this assemblage (Figure 13B) exhibits a large Ni/Fe ratio with composition close to that of pentlandite (Figure 14B). They are smaller in size and are only found in this assemblage. A small mineral with stoichiometry close to that of $Na_2S$ has been observed within dusty patches of DC16-309, although no diffraction or proper characterization could be made due to the loss of the DC16-309 FIB section after the first STEM session. From EDX spectra, the stoichiometry is closer to that of sulfide rather than that of sulfate.

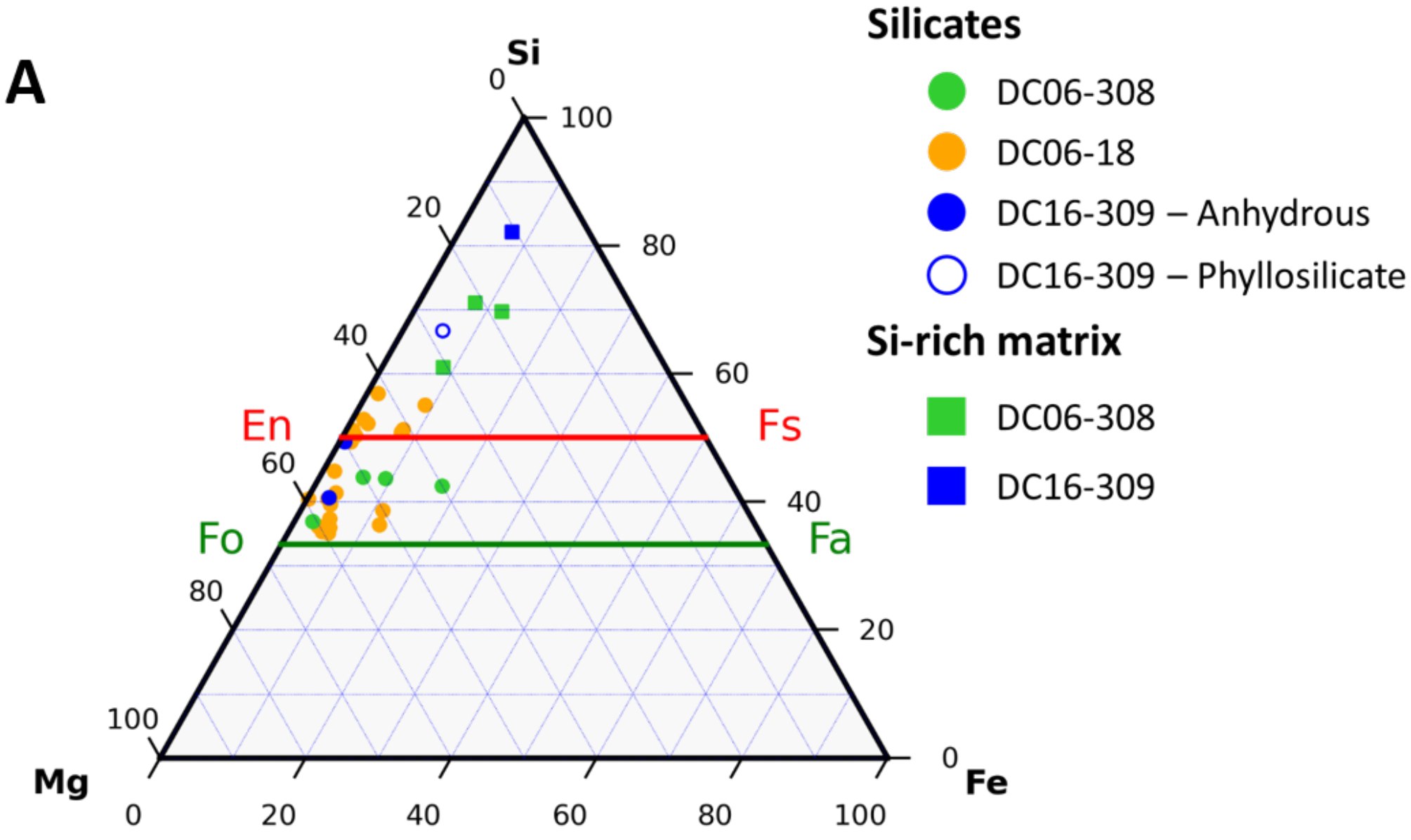

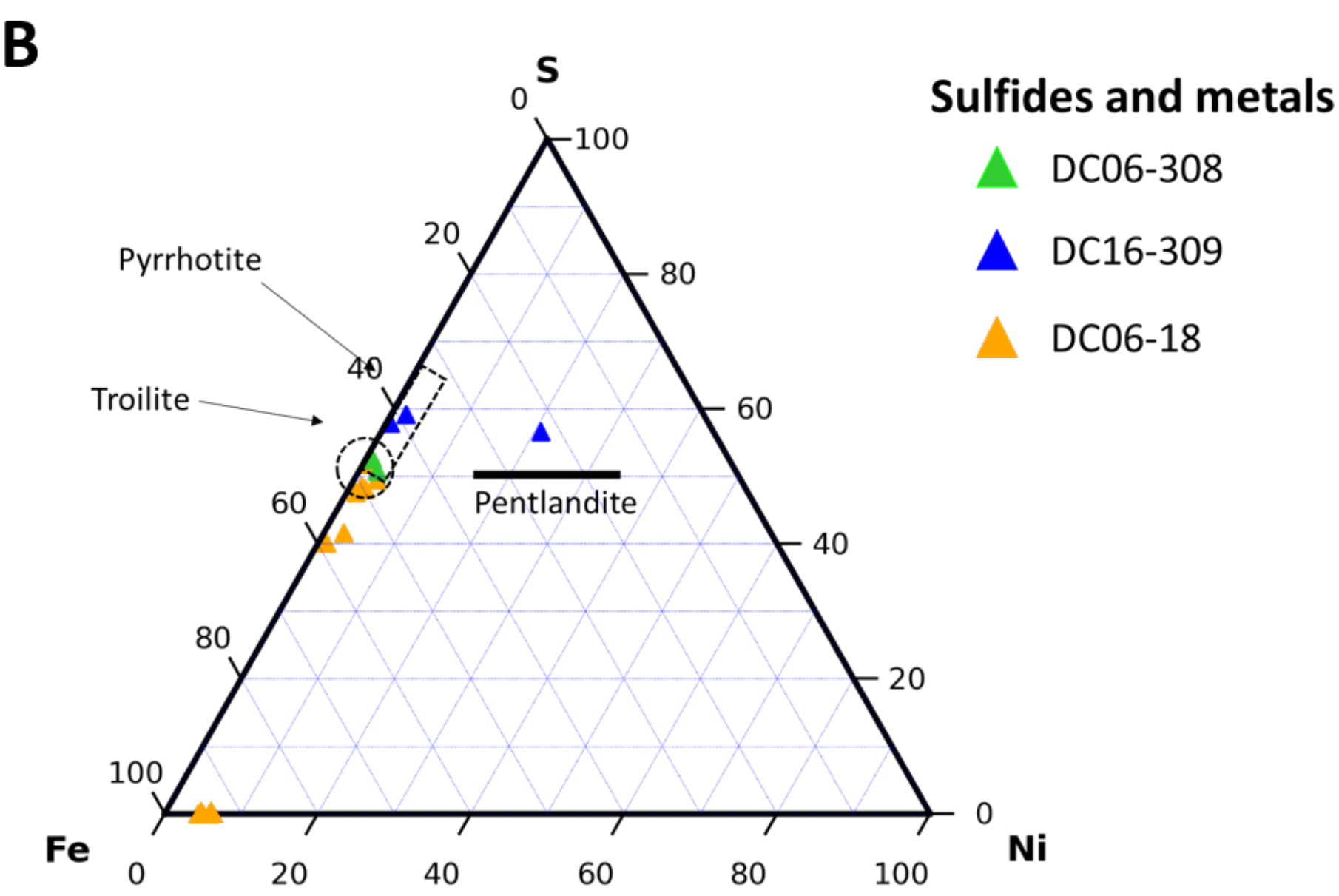

*Figure 14: (A) Ternary diagram displaying the composition (in at%) of silicate mineral and matrix in assemblages found in DC06-308, DC16-309 and DC06-18. (B) Ternary diagram displaying the composition (in at%) of Fe sulfides and metals in assemblages from UCAMMs DC06-308, DC16-309 and DC06-18.*

### 3.2.3 Mineralogy of UCAMM DC06-18

UCAMM DC 18 contains a region rich in minerals (Figure 15). These minerals are found in domains bounded by smooth organic matter. The individual domains are less than a micron in size, averaging around 300 nm. In this region, the porosity is important with pore sizes up to 1 µm. The largest domains are often crystalline and are assemblages of olivine, pyroxene and Fe-sulfide. Figure 16a and b respectively show two such assemblages for which the individual grains are on the order of 100-200 nm. The grain boundaries form 120° junctions attesting to a high temperature equilibrated texture. Pyroxenes are most frequently poor in calcium and contain abundant intergrowths of ortho and clinopyroxene. Some Ca-rich pyroxenes were also detected. One of these assemblages contains Fe-Ni metal (Figure 16c). A rim of Fe-sulfide suggesting an episode of sulfurization surrounds the metal grains. Many other domains also contain crystals but smaller in size (Figure 16d). In this case, an amorphous $SiO_2$-rich material cements them. For the crystalline silicates (olivine and pyroxene), whatever their grain size, the Mg / (Fe + Mg) ratio is between 0.9 and 1, with a dominant value of the order of 0.95. In a given domain, this ratio is constant from crystal to crystal, which shows that these assemblages are chemically balanced. Some domains, 100-200 nm in size, are dominated by amorphous silicate containing numerous opaque inclusions of nanometric size (Figure 16d). Their texture and size are comparable to that of GEMS that are frequently found in CP-IDPs. The inclusions are mainly Fe-sulfide. Others, rarer, carry iron, but we could not determine if these are Fe-metal or Fe-oxide. These phases will be quoted as "GEMS-like" phases. The bulk compositions of these GEMS-like domains have a fairly large chemical variability, which is illustrated in the Mg-Si-Fe, Fe-Mg-Si and Fe-Si-S ternary diagrams (Figure 17). The field of composition that emerges is comparable to that of GEMS in IDPs (Keller and Messenger, 2011). Note that the bulk compositions of the crystalline domains draw a comparable composition field.

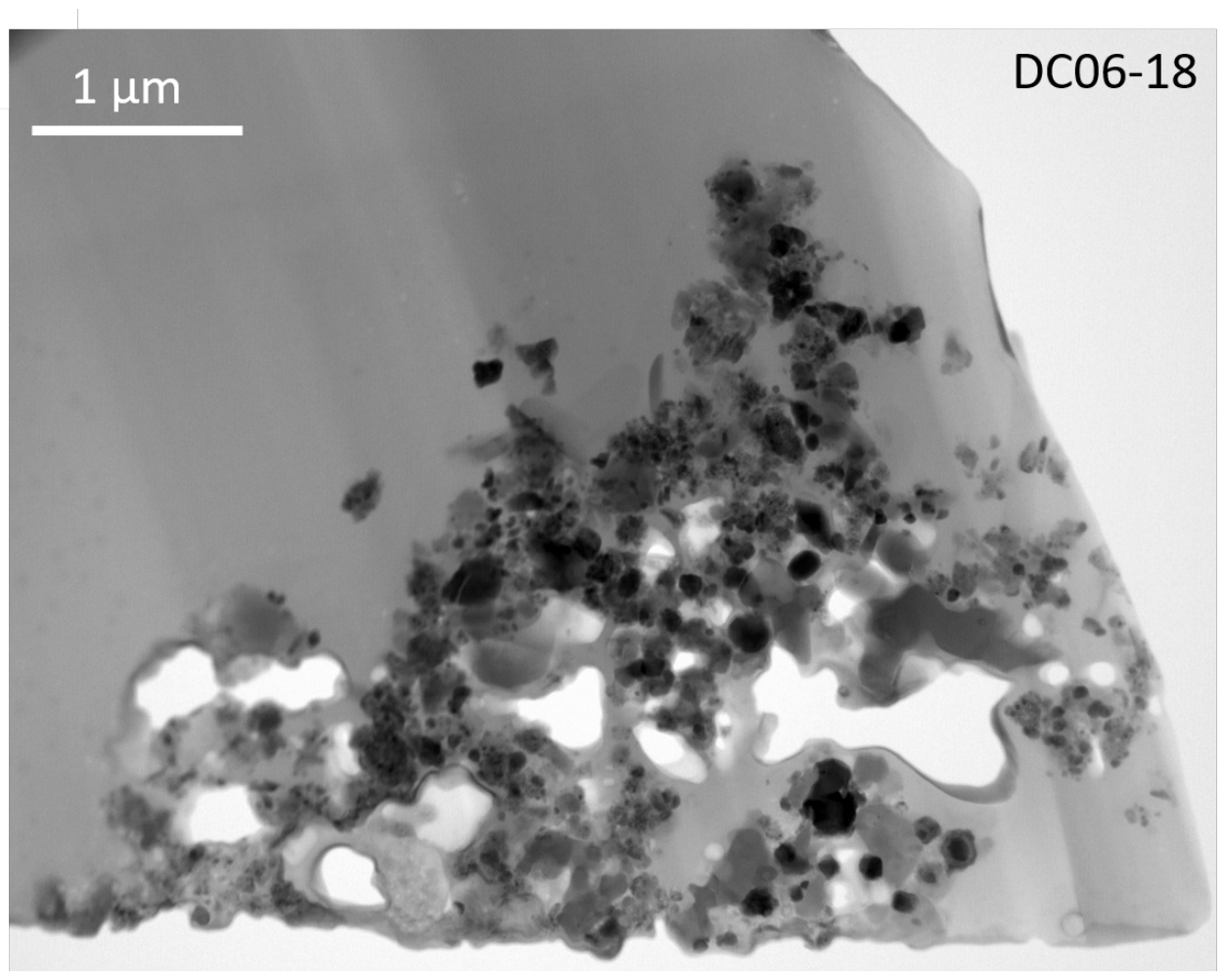

*Figure 15: General STEM bright field image of the mineral-rich area in DC06-18. Fe-Ni metal and Fe-sulfides appear in dark, silicates in medium grey, organic matter in light grey and pores in white.*

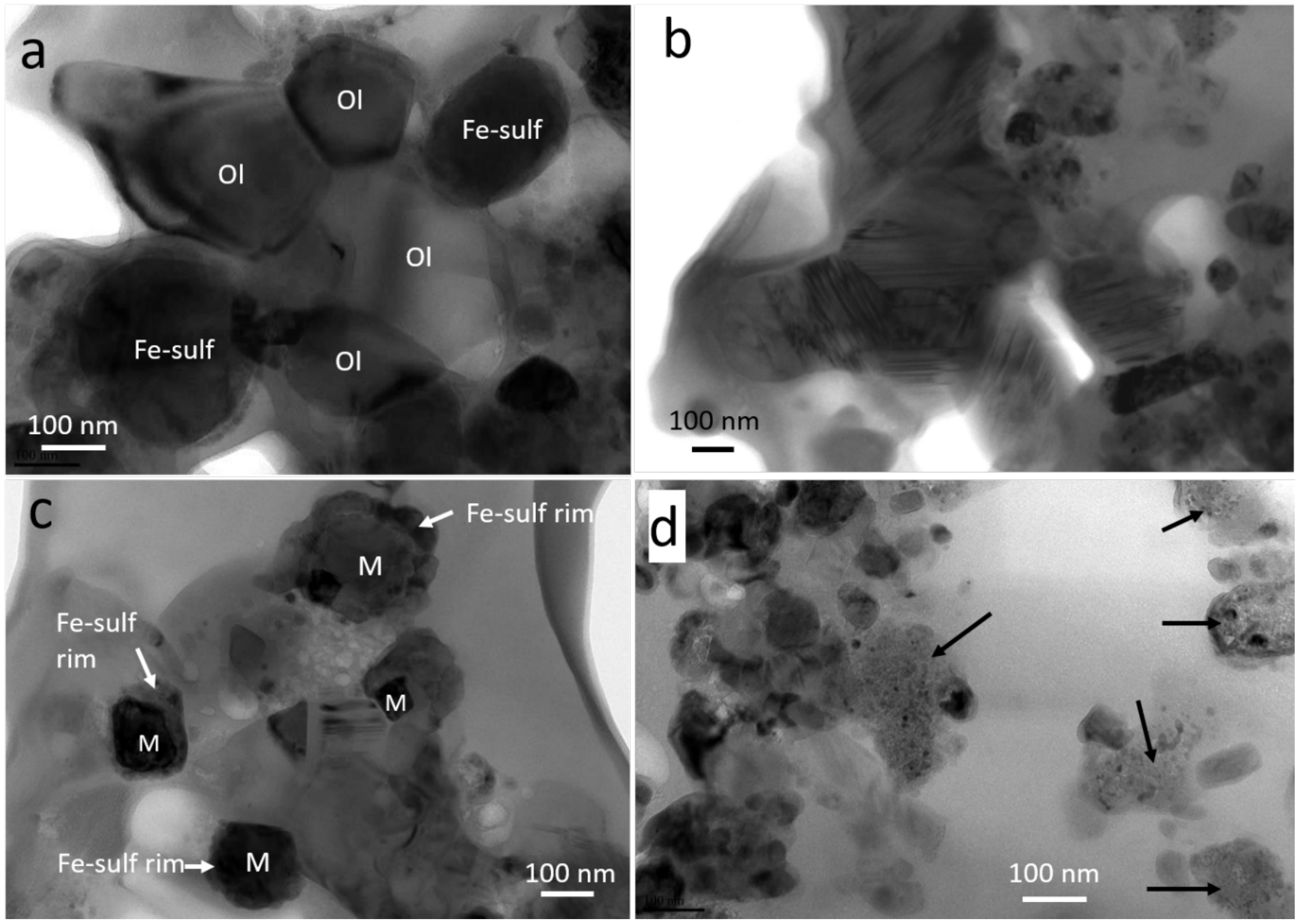

*Figure 16: Mineralogy of the region rich in minerals in UCAMM DC06-18 characterized by bright field TEM. (a) Domain dominated by coarse-grained olivine (Ol) and Fe-sulfide (Fe-sulf) grains with an equilibrated texture. (b) This domain consists of Fe-Ni metal (M) having a rim of Fe-sulfide (Fe-sulf rim). The bottom-middle area (medium grey) is an assemblage of pyroxene and olivine. Note the presence of organic matter (light grey) containing bubble, which contrast with the smooth surrounded organic matter. (c) Assemblage of coarse-grained enstatite with an equilibrated texture. Note the presence of ortho-clino intergrowth within the grains that are at the origin of the internal striation. (d) The right part consists of a fine-grained assemblage of crystalline silicates cemented by a $SiO_2$-rich amorphous material. GEMS-like domains are arrowed.*

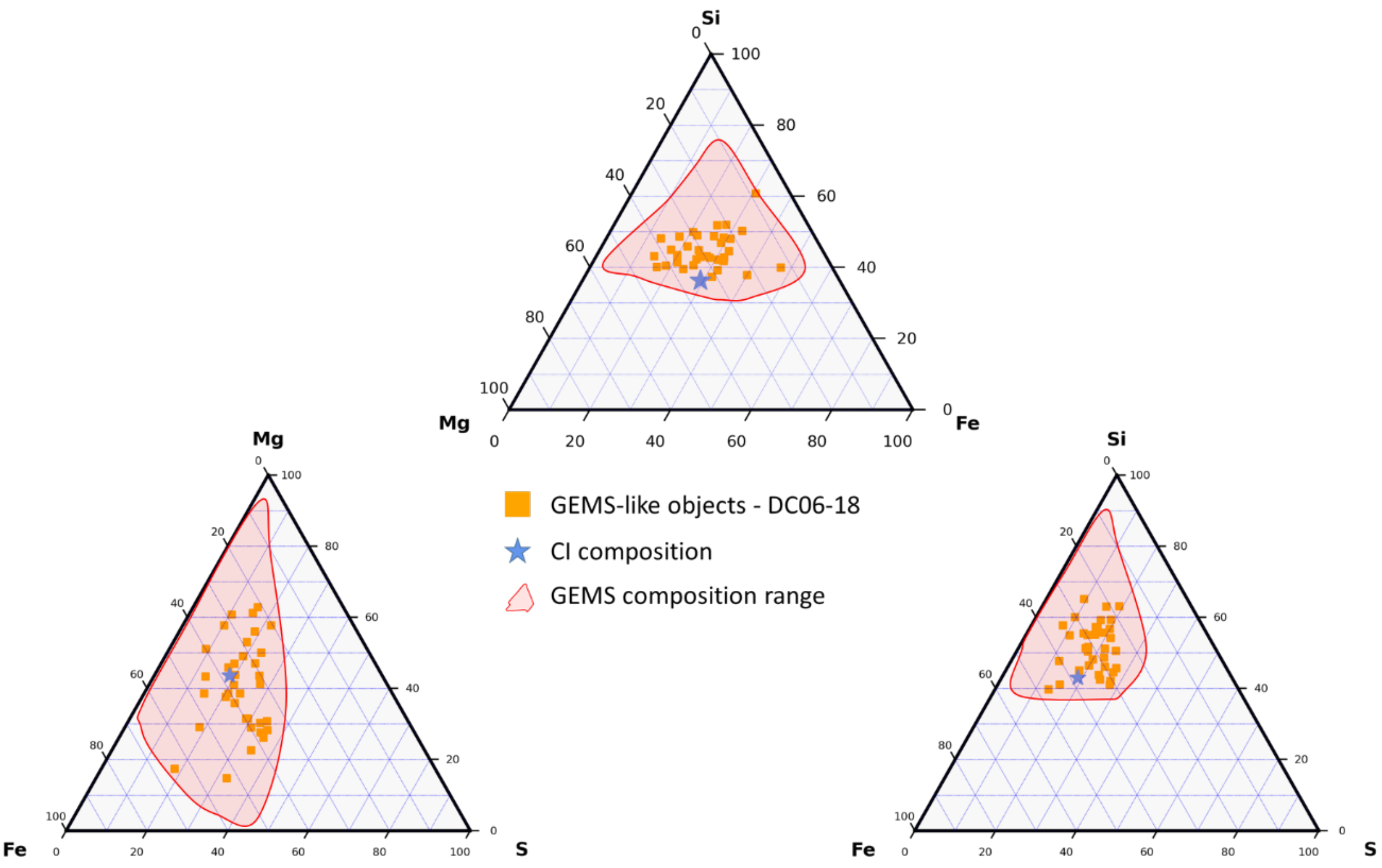

*Figure 17: Mg-Fe-Si ternary composition field of the GEMS-like objects (in at%). Each point corresponds to the bulk analysis of the domains which are typically 100- 200 nm in diameter. The range of GEMS compositions is displayed by the shaded red area. GEMS Data from Keller & Messenger (2012; 2011) and Messenger & Keller (2015). CI data from Lodders (2010).*

# 4 Discussion

## 4.1 Comparison of the different types of organic matter in extraterrestrial matter.

### 4.1.1 Elemental ratios and molecular signatures of cometary organic matter

The Rosetta/COSIMA mass spectra of cometary dust particles from 67P/C-G inferred that 67P/C-G particles were made of up to 50 wt% of organic matter. This ratio is in the range of what is found for UCAMMs, which display very large organic/mineral ratio (up to 90 vol% of carbon). The solid organic component of comet 67P/Churyumov-Gerasimenko characterized by Rosetta/COSIMA exhibits similarities with IOMs extracted from carbonaceous chondrites (Fray et al., 2016), although the H/C measurement in 67P/Churyumov-Gerasimenko particles (H/C=1.04±0.16) show that organics were more saturated that organic material in these IOMs (Isnard et al., 2018). This suggests that 67P/Churyumov-Gerasimenko organic material is more pristine than the IOMs extracted from carbonaceous chondrites.

For comet 81P/Wild 2, Stardust analysis revealed on the contrary a low carbon content (Brownlee, 2014). This can partly be explained by the collection method of the Stardust mission, which probably volatilized a large part of the organic species during the impact with the aerogel.

Matrajt et al. (2008) identified rare carbonaceous phases in Stardust samples, with an amorphous nature and approximate solar C and D/H isotopic compositions. Among the rare carbon-rich phases detected in Stardust samples, De Gregorio et al. (2017) also revealed nanometric Cr-rich magnetite coated with conformal layers of poorly graphitized carbon (PGC). PGC layers formation is consistent with Fischer-Tropsch-types formation process, occurring on primary carbide condensates or metal grains. This process is consistent with nebular oxidation of reduced material, showing additional evidence of large-scale radial mixing in the early solar system, where high-temperature material could have been brought out to the outward solar system.

The N/C elemental ratio of UCAMMs reveals relatively high nitrogen content compared to carbon. The STXM-XANES analyses of UCAMMs revealed the presence of three types of organic matter with different nitrogen compositions (Figure 18) and two of them are often closely associated to each other, as type II OM only appears embedded in with type I OM. Atomic N/C ratios are variable in type III OM, with values ranging from 0.08 for mildly heated particles (such as DC16-30, see section 4.3) to 0.2, while type I and II N/C ratios range from below detection limit to 0.05. Type III OM being the main N carrier in the UCAMM organic matter, the bulk N/C atomic ratios reported in Table 3 for each UCAMM should be lower than the N/C values determined for their type III OM. This is observed in most samples, except for DC06-18 and DC16-30 (see Table 3). As the N/C atomic ratios were measured on different UCAMM fragments, this suggests heterogeneity of N/C atomic ratios in UCAMMs. The elemental composition of type I and II OMs indicate a N/C atomic ratio similar to that of IOM extracted from CCs ($0<N/C<0.05$). The N/C atomic ratio of 67P/C-G cometary dust is consistent with previous analysis in comets (81P/Wild 2, 1P/Halley) with values ranging between 0.02 and 0.06 (Fray et al., 2017a) similar to that measured in type I OM in UCAMMs and in the IOM from carbonaceous chondrites. The low N/C atomic ratio of Stardust organics is probably biased by the collection methods, which implied high temperature heating of the samples during impact with the aerogel. This suggests that the measured Stardust N/C atomic ratio measured can be seen as a lower limit of the original nitrogen content of the comet organic matter. The different N contents of the three OMs in UCAMMs suggest different origins for type I/II OM and type III OM. UCAMMs' type I and II OMs embed high temperature crystalline minerals such as pyroxene and olivine. The association of high temperature phases with types I and II OMs possibly indicates an inner solar system origin although the mechanism of formation or accretion of IOM-like material is still debated (Alexander et al., 2017; Alexander et al., 2007; Gourier et al., 2008; Remusat et al., 2006). However, the IOM-like phases observed in UCAMMs extends over larger areas that the organic phases found in meteorites. Despite their similarities with IOM, an additional process is needed to account for the spatial extent of type I and II organic. This large organic matrix could result from fluid-interaction during a first parent body processing. The presence of two different organic phases (type I and II) can also result from several scenarios. The first hypothesis consists in the accretion of originally heterogeneous organic precursors that evolved into type I and II organics. Another possibility would be a redistribution of the organic precursors through fluid-interaction during parent body processing, to form two distinct phases {Le Guillou, 2014 #3144}. Finally, the identification of Na-rich nanophases located within the patches type II OM could also explain the difference observed in the STXM-XANES spectra between type I and II OM.

The type III OM presents much higher N/C atomic ratios with values up to 0.2. In addition, it harbors a smooth texture, is devoid of crystalline minerals and can extend over several $\mu m^2$ in area. GEMs and composite assemblages similar to GEMS can sometimes be found associated with type III OM in some samples (DC06-09-19 from Dobrica et al. 2012, DC06-43, DC06-18). The production of such a N-rich organic matter from ices' irradiation by galactic cosmic rays (GCR) can be achieved at large heliocentric distances and could be one formation mechanism of type III OM (Augé et al., 2016; Dartois et al., 2013). N-rich volatiles can be retained by large enough bodies beyond a so-called "nitrogen snow line" in the form of ices. Experimental high energy ion irradiations on accelerators of ice mixtures representative of solar system icy body surfaces can indeed produce residues showing infrared features comparable to the ones recorded in the organic matter spectra of UCAMMs (Augé et al., 2019; Augé et al., 2016; Dartois et al., 2018; Rojas et al., 2020).

Recent analysis of the PanSTARRs comet based on $N_2$/CO measurement in the comet's coma indicates that its organic content could exhibit an elemental N/C atomic ratio around 0.15 (Biver et al., 2018), a value consistent with what is found in the type III OM of UCAMMs. However, this ratio for the PanSTARRs comet is derived from the CO and $N_2$ species, and the calculated N/C ratio might not be directly comparable with the OM in UCAMMs, which contain generally less oxygen that CCs IOM (Dartois et al., 2018).

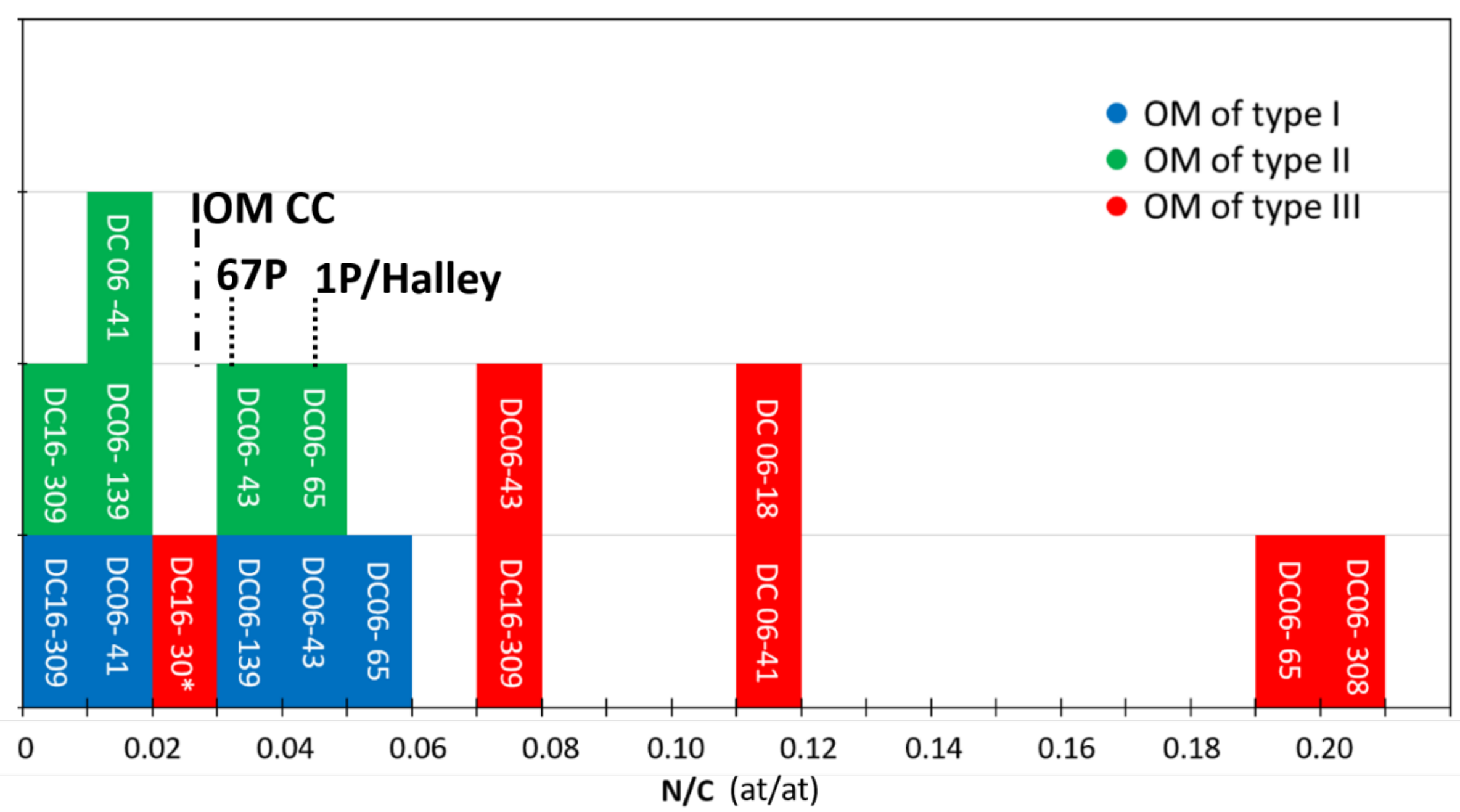

*Figure 18: Histogram of the distribution of N/C atomic ratios for the 3 different organic matter types found in each UCAMM presented in this study. The upper values of N/C atomic ratio measured in carbonaceous chondrites IOM and cometary material are also reported. *The low N/C atomic ratio for type III OM in DC16-30 could be explained by a higher heating during atmospheric entry, see section 0.*

*Table 3 : Bulk N/C atomic ratios for five UCAMMs measured by electron microprobe, and N/C atomic ratios for their respective type III organic matter measured by STXM-XANES.*

| | Bulk N/C (EMPA)* | Type III N/C (STXM-XANES) |
|---|---|---|
| DC06-18 | 0.17 ± 0.04 | 0.11 ± 0.02 |
| DC06-41 | 0.06 ± 0.02 | 0.12 ± 0.02 |
| DC06-65 | 0.15 ± 0.04 | 0.20 ± 0.03 |
| DC06-43 | 0.04 ± 0.03 | 0.08 ± 0.03 |
| DC16-30 | 0.09 ± 0.06 | 0.03 ± 0.02 |

*from Dartois et al. (2018)

Analysis of Stardust samples allowed STXM-XANES characterization of different organic fragments of 81P/Wild 2 cometary dust. Among the samples presenting carbonaceous material, track 80 is thought to be the most pristine and more representative sample of indigenous cometary organics from Stardust collection (De Gregorio et al., 2011). It appears that organic matter of Stardust presents similar carbon speciation as type I in UCAMMs (see Figure 19). Both spectra display their main first peak at 284.8 eV, corresponding to the alkene carbon while type II is more aromatic (the first peak is shifted toward the 285 eV value). These spectra are also close to that of insoluble organic matter of carbonaceous chondrites (Alexander et al., 2017; Changela et al., 2018; Le Guillou et al., 2014; Vinogradoff et al., 2018). Type I OM of UCAMMS also display a very similar spectrum compared to the spectrum of carbonaceous chondrite IOM from QUE99177. The main peaks at 284.8 eV and 286.4 eV are also present and some carboxylic absorption is also visible in the same trend as for Stardust.

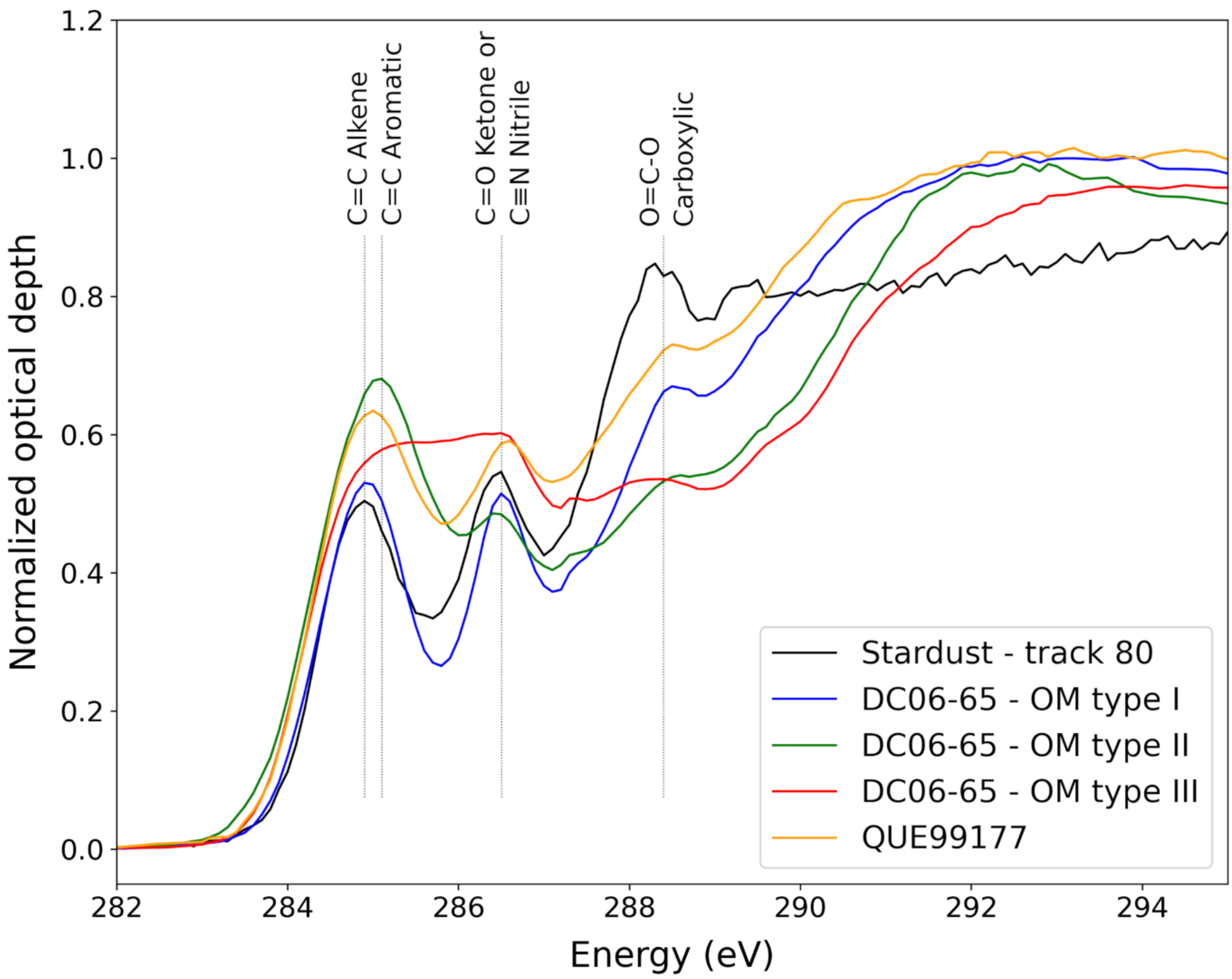

*Figure 19: STXM-XANES spectra of the 3 different types of OM in UCAMM DC06-65 compared to Stardust organic pristine particle from track 80 and IOM from QUE99177.*

### 4.1.2 Characterization of the organic-mineral interface

In the STXM-XANES spectra at the C-K edge, type II OM does not display the same drop in absorption between 285 eV and 286.5 eV as for type I OM. The presence of inorganic phases within the dusty patches in type II OM could explain this continuous absorption. This may possibly be explained by C=C-R configuration, with aromatic carbon linked to other carbon or heteroatoms such as N or O instead of hydrogen, shifting the absorption energy of aromatic carbon toward higher values (>285 eV).

GEMS-like inclusions can be present in type I or type III OM. These fine-grained phases could in turn contain organic matter, like what is observed in DC06-18 (see Figure 9). In that case, the GEMS-like inclusions are enclosed in type III OM and the N content of the organic matter that is located inside the inclusion can reach N/C atomic ratios up to ~0.5. This value is higher than the maximum atomic N/C atomic ratio observed for type III OM, which is around 0.2 (Table 2). Regarding the organic residue produced by irradiation, the samples were slowly heated,up to 300°C, in order to reproduce the IR signature of UCAMMs. This suggests a mild thermal alteration of the precursor of type III OM in UCAMMs, maybe during atmospheric entry (Augé et al., 2016). In the case of DC06-18, the organic matter trapped inside the GEMS-like inclusion could have been protected from such a mild thermal event, and could thus have better preserved the initial N/C atomic ratio of the N-rich organic matter.

UCAMMs show a bulk Na enrichment compared to CI. The high bulk Na abundances of UCAMMs, and the high N/C ratio of their organic matter are exceptional in the sense that they are not observed in any other kind of extraterrestrial material analyzed in the laboratory. Na is common in comets, as evidenced by the observation of a sodium tail in comet Hale-Bopp (Cremonese et al., 1997), and of a systematic Na enrichment in particles collected and analyzed by COSIMA during the Rosetta mission around comet 67P/Churyumov-Gerasimenko (Bardyn et al., 2017). The host phase of Na in comets has however not been identified yet. Figure 9 and Figure 10 show an increase of the Na concentration close to GEMS-like phases in DC06-18, which appears sometimes correlated with an increase of the N concentration in the organic matter. The Na/Si atomic ratio is about twice that of CI in GEMS-like phases in DC06-18. The partial correlation of the Na and N abundances suggests the presence of at least two Na-bearing phases, one of them being correlated with the N content of the organic matter close to the inorganic phases. Alternatively, the Na halo observed in Figure 9 and Figure 10 around the mineral assemblages could possibly be explained by a mild heating event inducing diffusion toward the organic matter of Na from the GEMS-like phase, with an atomic Na/Si ratio ~ 2x CI, toward the organic matter.

The N/C atomic ratio in comets is lower than that of the Sun, which suggests the presence of a hidden nitrogen reservoir in comets (Biver et al., 2018; De Gregorio et al., 2011; Fray et al., 2017b; Jessberger et al., 1988; Lodders, 2010)}. Semi volatile ammonium salts were recently identified in comet 67P/Churyumov-Gerasimenko, and could represent a significant fraction of the nitrogen present in comets (Altwegg et al., 2020; Poch et al., 2020). To match the reflectance spectra of 67/C-G, these ammonium salts are associated with a finely grinded Fe sulfide (Poch et al., 2020). The observed N and Na partial correlation in UCAMM DC06-18, and the identification of a mineral in type II OM with a composition close to that of $Na_2S$ questions the relevance of testing $Na_2S$ in a combination with ammonium salts to reproduce the reflectance spectra of 67P/ Churyumov-Gerasimenko.

## 4.2 Comparative mineralogy of cometary dust particles

It has been estimated that Earth accretes around 30 000±20 000 tons of sub-millimetric extraterrestrial material every year. With 20% to 50% of this flux reaching Earth surface, recent study showed that $5200^{+1500}_{-1200}$ tons of extraterrestrial particles falls on Earth every year (Love and Brownlee, 1993; Rojas et al., 2021). Dynamic models predict that most of this cosmic dust originates from active Jupiter-family comets (Nesvorný et al., 2010; Plane, 2012). CP-IDPs (Bradley, 2016; Ishii et al., 2008) and UCAMMs are the two cosmic dust populations that show the strongest evidence to originate from comets (Dartois et al., 2013; Dartois et al., 2018; Duprat et al., 2010; Nakamura et al., 2005). These samples can be analyzed in laboratory, and the results directly compared with the analyses from space missions that investigated comets 1P/Halley, 9P/Tempel1, 81P/Wild2 and 67P/Churyumov-Gerasimenko (Giotto, Vega, Deep Impact, Stardust and Rosetta missions).

The bulk composition of the hypocrystalline-like assemblage in DC06-308 is CI-like, see also Dobrică et al. (2012) for other examples of such assemblages. The bulk composition of the other assemblages studied here is enriched in Si compared to CI (Figure 20). Si-rich compositions are also found in primitive phases in primitive meteorites (Le Guillou and Brearley, 2014; Le Guillou et al., 2015; Le Guillou et al., 2019; Leroux et al., 2015; Vollmer et al., 2020; Zanetta et al., 2021) in GEMS in CP-IDPs (Keller and Messenger, 2011) and in comet 67P/Churyumov-Gerasimenko (Bardyn et al., 2018).

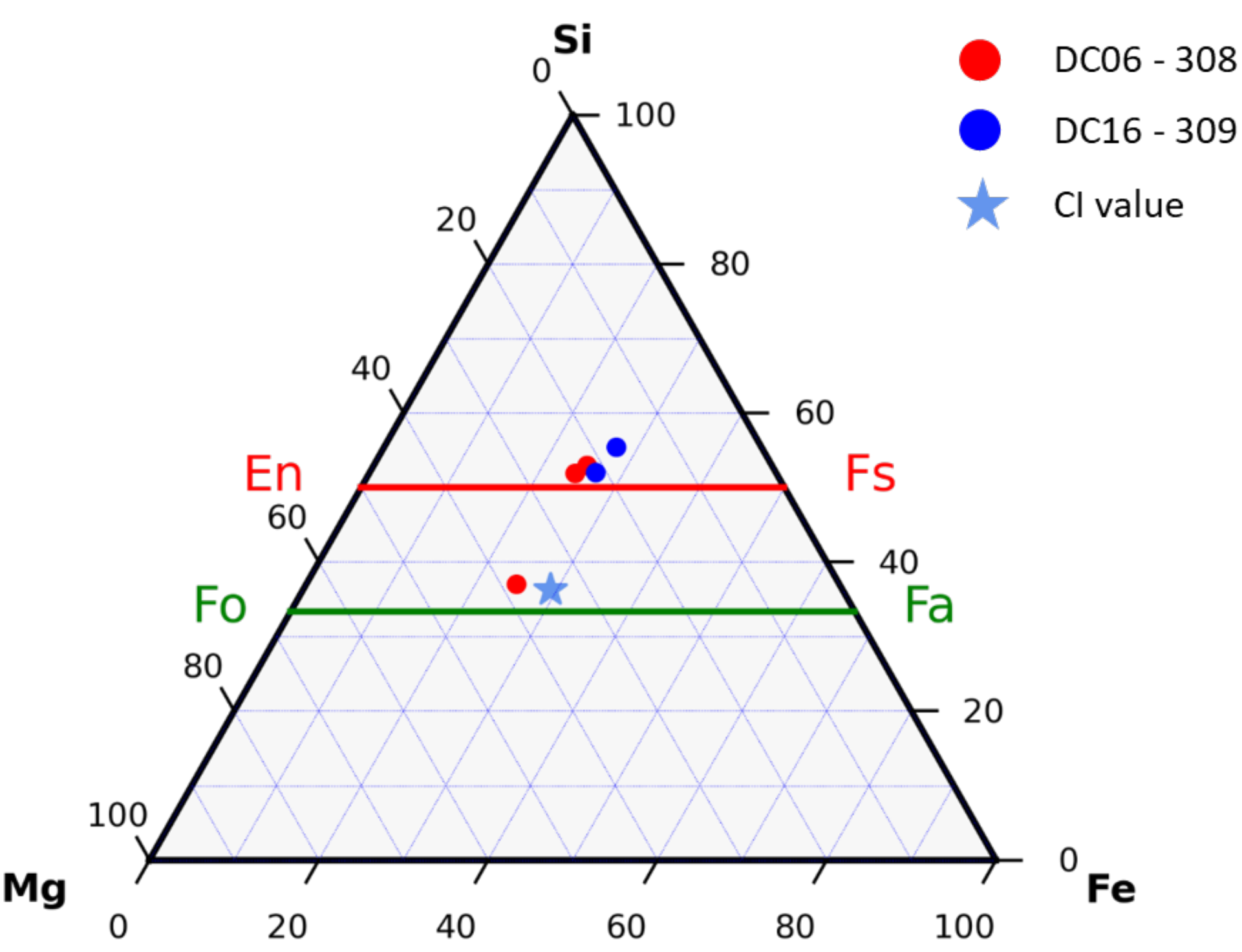

*Figure 20: Ternary diagram showing the bulk composition (in at%) of the different assemblages presented in this study. The blue star represents the CI value from Lodders (2010).*

The individual mineral phases in the 11 UCAMMs analyzed so far (this work, (Dobrică et al., 2012) mostly consist in small mineral assemblages and scarcer isolated minerals embedded in an abundant carbonaceous matrix (excepted for DC06-43 for which the analyzed fragment contain a 10µm-sized aggregate of minerals). Whereas the analyses of UCAMMs DC06-308, DC16-309 and DC06-18 revealed mineral compositions in agreement with previous observations (Charon et al., 2017; Dobrică et al., 2012; Engrand et al., 2015), one phyllosilicate-like phase was unexpectedly found in DC16-309 (Figure 13A).

Considering previous work, in addition to the present study, the mineralogy of UCAMMs is summarized in Figure 21. Crystalline minerals from this study consist of low-Ca Mg-rich pyroxenes (with stoichiometry ranging between $En_{60}$ and $En_{97}$) and Mg-rich olivines (stoichiometry comprised between $Fo_{75}$ and $Fo_{99}$) with rare Ca-rich pyroxenes and Fe(Ni) sulfides. Fe sulfides are often decorating silicate mineral with composition matching that of troilite and/or pyrrhotite (low-Ni Fe sulfides) but occasionally pentlandite crystals have been observed. Glassy phases are also observed in 3 UCAMMs (Charon et al., 2017; Dobrică et al., 2012; Engrand et al., 2015) that resemble GEMS found in primitive IDPs (Keller and Messenger, 2011). Secondary minerals have also been observed in UCAMMs (Dobrică et al., 2012) such as small iron oxides, carbonates and sphalerite. Small Na-rich crystals with stoichiometry close to $Na_2S$ have been observed located within the dusty patches in DC16-309. A mineral with a fibrous texture reminiscent of that of a phyllosilicate has been observed in DC16-309 (see Figure 13A).

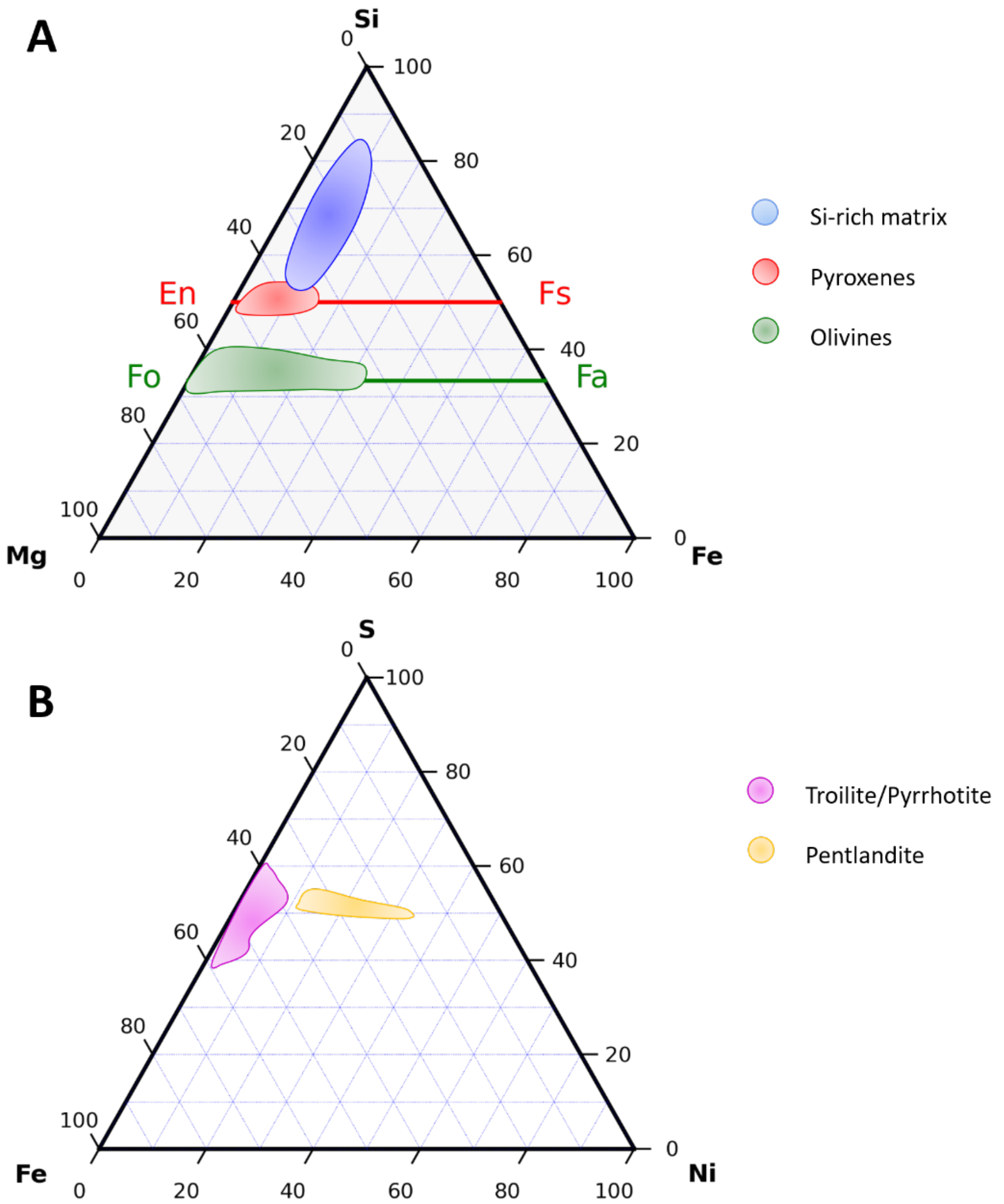

*Figure 21: (A) Ternary diagram displaying the composition ranges (in at%) of silicate phases in UCAMMs DC06-308, DC16-309, DC06-18 along with previous silicate composition from UCAMMs DC06-19 and DC06-41 (B) Ternary diagram displaying the composition ranges (in at%) of sulfur phases in UCAMMs DC06-308, DC16-309, DC06-18 along with previous composition from UCAMMs DC06-19 and DC06-41. Additional data from (Dobrică et al., 2012)*

Although no mineralogical characterization was possible with the instruments on board the Rosetta mission, but a bulk Si-rich composition compared to the Solar composition was observed in 67P/C-G dust particles, with Mg/Si and Fe/Si atomic ratios of 0.11 ± 0.03 and 0.29 ± 0.1, respectively (Bardyn et al., 2018). These values are compatible with that of the Si-rich matrix measured in the UCAMM mineral assemblages in DC06-18, DC06-208 and DC16-309 (Figure 14).

Compared to CP-IDPs, UCAMMs mineralogy show similar main mineral components (Mg-rich crystalline silicate, FeS and GEMS) and accessory phases alike what is observed in CP-IDPs (Dobrică et al., 2012; Ishii et al., 2008; Levasseur-Regourd et al., 2018). In addition to metal-poor GEMS found in UCAMM DC06-09-19 (Dobrica et al. 2012), GEMS-like objects have been found in UCAMMs DC06-18 and DC06-43 among the 4 UCAMM-bearing minerals investigated in this study. The GEMS-like inclusions are either contained in type I phase or type III. Si-rich glass and GEMS-like objects present in UCAMMs only contain rare Fe metal inclusions compared to what is observed in CP-IDPs. Overall, the occurrence of pyroxene and olivine mineral grains lead to pyroxene/olivine ratio equal or larger than 1 for UCAMMs (Dobrică et al., 2012). This pyroxene-dominated mineralogy is also seen in CP anhydrous IDPs (Donahue, 1999).

Direct measurements of Halley comet during Vega-1, Vega-2 and Giotto missions brought the first insight into the cometary dust composition (Jessberger et al., 1988; Kissel et al., 1986; Kissel and Krueger, 1987). Analysis of the mass spectra of Halley particles revealed the lack of Fe-rich pyroxene and olivine and the survival of Mg-rich silicate, a mineralogy close to that observed in UCAMMs (Jessberger et al., 1988). The large occurrence of pyroxenes is also seen in Stardust samples (Zolensky et al., 2006) although olivines with wide compositional range are also common in Stardust tracks (Frank et al., 2014). Spectral analysis of comet Hale-Bopp also revealed a Mg-rich silicate mineralogy, containing both olivine and pyroxenes (Malfait, 1999 ; Wooden et al., 2000; Wooden et al., 1999). Spectral analysis of the comet 9P/Tempel 1 during Deep Impact mission exposed a mineralogy dominated by Mg-rich silicates with a pyroxene/olivine ratio close to unity along with the presence of carbonate and potential phyllosilicate (Lisse et al., 2006). Mineralogical zoning in disks have been observed (Bouwman et al., 2008; van Boekel et al., 2005), but no systematic trend with radial distance to the star has been described yet. As both CP-IDPs and UCAMMs are thought to originate from comets, a Px/Ol ratio close to 1 may be an indication of cometary origin. With high spatial resolution in the observation of disks, ground-based interferometers such as MATISSE, and the soon-to-be-launched James Webb Space Telescope, will give new insight into mineralogical zoning in protoplanetary disks.

UCAMM DC06-308 presents a mineral assemblage exhibiting an igneous-like texture (Figure 12) similar to what was observed in Dobrică et al. (2012), and reminiscent of chondrule fragments observed in Stardust samples (Nakamura et al., 2008). It resembles a chondrule-like object, with porphyritic olivines embedded in a Si-rich matrix. This assemblage could have formed through a chondrule-forming event where dust particles were melted, until crystallization of mineral phases with equilibrated compositions. This event could not have happened after the incorporation of the minerals in the organic matter of UCAMMs (i.e. typically not during atmospheric entry), as it would have required temperatures above 1200°C which would have strongly modified or destroyed this organic matter. The presence of such high-temperature assemblages within cometary dust is compatible with the presence of crystalline silicates and points toward radial mixing in the protosolar disk where high temperature material were transported outward to the outer solar system (Birnstiel et al., 2012; Bockelée-Morvan et al., 2002; Ciesla, 2009).

Secondary minerals such as carbonate (Dobrică et al., 2012), and the phyllosilicate candidate found in this study are also observed in UCAMMs (see Figure 13 and Figure 22). Spectral observation of the ejecta subsequently to Deep Impact mission on comet 9P/Tempel 1 was interpreted to contain about 5% of carbonates and up to 9% of phyllosilicate (Lisse et al., 2007a; Lisse et al., 2006), although this remains controversial (Harker et al., 2005). Wooden et al. (2004) and Lisse et al. (2007b) also claimed that a phyllosilicate signature in comet Hale-Bopp could not be ruled out. However, definitive in-situ detection of cometary phyllosilicate has not been achieved yet. Analyses of Fe-sulfides in Stardust tracks revealed the presence of pyrrhotite, pentlandite and orthorhombic cubanite (Berger et al., 2011). The combination of these sulfides points toward aqueous alteration at low temperature (<210°C). This aqueous alteration could have happened on an asteroidal parent body, requiring a later transport mechanism of the altered components to the comet-forming region. Alternatively, aqueous alteration on the comet could be considered (see paragraph 4.4).

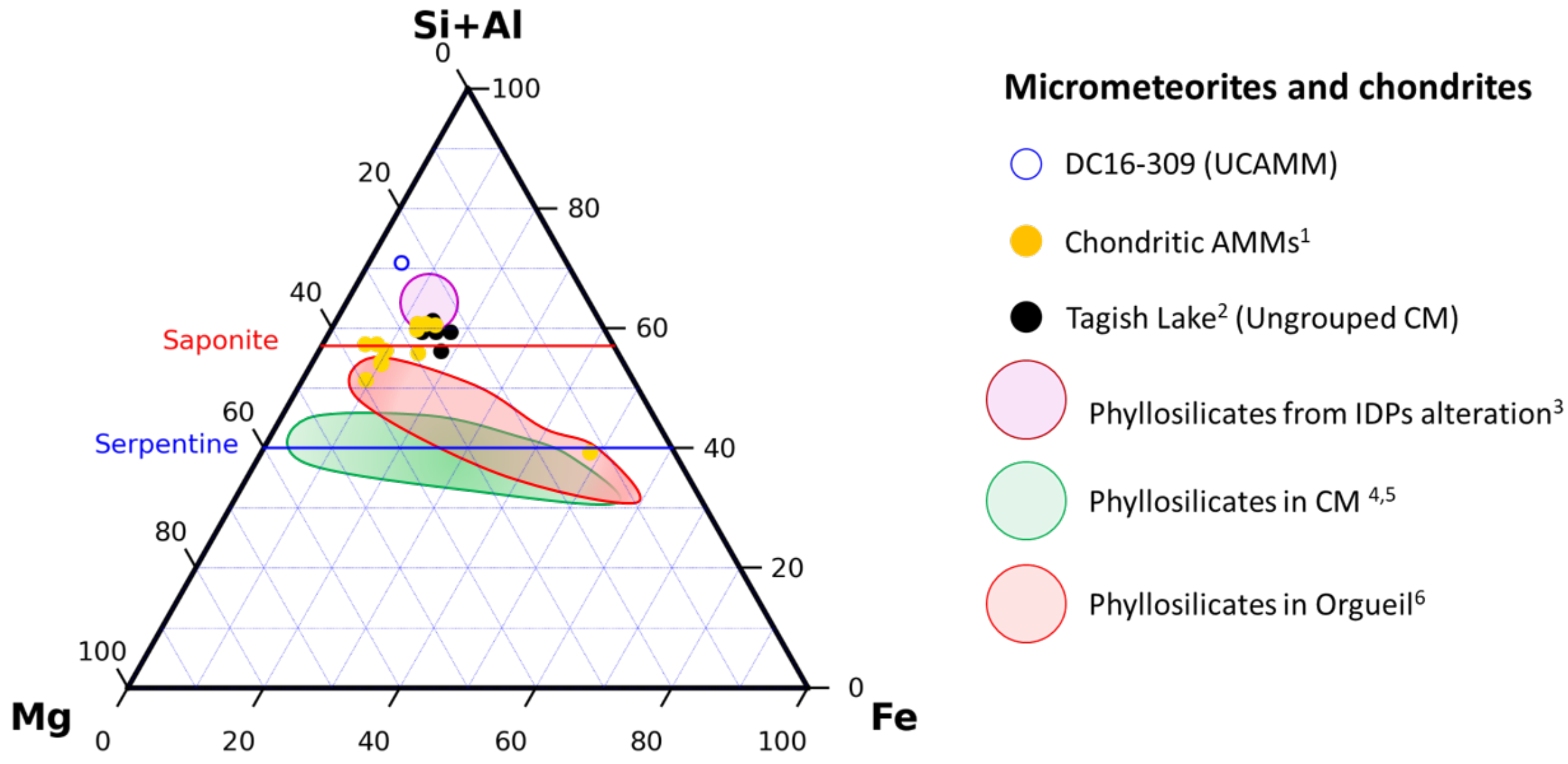

*Figure 22: Ternary diagram displaying the composition (in at%) of the phyllosilicate-like mineral in DC16-309 (open blue circle) along with that of phyllosilicates from carbonaceous chondrites (CCs), Antarctic micrometeorites [1] data from Dobrica et al (2018) and Noguchi et al. (2002), and aqueously altered GEMS (Nakamura-Messenger et al., 2011). Data for CCs from [6](Tomeoka and Buseck, 1988), [4](Bunch and Chang, 1980), [5](Browning et al., 1996), [2](Noguchi et al., 2002).*

The composition of the phyllosilicate-like mineral in DC16-309 exhibits an unusual composition with a high Al/Si ratio of 0.22. It does not match that of saponite or serpentine stoichiometry and is Al-rich and Fe-poor compared to phyllosilicates in CCs. Hydrous silicates exhibiting compositions similar to the fibrous mineral observed in DC16-309 are found in Tagish Lake meteorite. Phyllosilicates obtained by aqueous alteration experiments of GEMS reported by Nakamura-Messenger et al. (2011) also display composition rather close than the fibrous mineral observed in UCAMM DC16-309. The alteration experiments were conducted in conditions simulating cometary surface alteration, suggesting the possibility of in-situ aqueous alteration.

Overall, UCAMMs displays rather scarce inorganic phases that probably originate from different locations and/or processes in the protoplanetary disk. High temperature crystalline phases formed first in the early solar system. Some of them might have experiment subsequent aqueous alteration on an asteroidal parent body where secondary minerals such as phyllosilicate and sulfides, would have formed.

The 3.2 µm band of the 67P/Churyumov-Gerasimenko nucleus is compatible with the presence of ammonium ($NH_4^+$), which constitutes a good spectral fit to the spectra (Poch et al., 2020), and would explain the nitrogen depletion in the coma (Altwegg et al., 2020). Among matching candidates, ammonium sulfate ($(NH_4^+)_2SO_4^{2-}$) has been proposed. The presence of small $Na_2S$ –like minerals embedded in dusty patches UCAMM DC16-309 could be related to the potential presence of this ammonium sulfate in comets.

## 4.3 Atmospheric entry vs. parent body heating

Since UCAMMs are collected on Earth after atmospheric entry, one has to decipher the effect of heating during atmospheric entry from possible heating processes on their parent body. As UCAMMs can reach sizes from tens to several hundreds of µm, they are expected to experience higher temperature rises than smaller particles such as IDP (tens of µm) during atmospheric entry. According to Love et al. (1994), MMs with sizes larger than 50 µm can reach temperatures above 800 °C during atmospheric entry heating.

The UCAMMs organic matter characteristics gives an upper limit to the extent of heating (<500°C) they could have experienced during atmospheric entry. Raman analysis of the organic matter of several UCAMMs (Dartois et al., 2018) display large D and G FWHMs bands (above 140 $cm^{-1}$). FWHM values are sensitive to the temperature indicating atmospheric entry heating, and larger bands corresponds to materials that experienced rather low heating temperature. The large D and G Raman bands in the UCAMMs spectra suggest they experienced temperature lesser than about 500°C, based on simulations from Bonnet et al. (2015). The presence of a nitrile (2220 $cm^{-1}$) peak in the infrared spectra of some UCAMMs (Dartois et al., 2018) also constrains temperature to values lower than about 500°C during atmospheric entry, as the nitrile content would not be preserved at higher temperature. It is also important to note that the measured N/C atomic ratio in UCAMMs can be considered as a lower limit value, as even a limited heating would reduce this ratio. The low nitrogen content of UCAMM DC16-30 type III OM (N/C = 0.03 ± 0.02, see Figure 18) could result from a more extensive heating during atmospheric entry. Indeed, as seen in Figure 1, DC16-30 displays a vesicular texture that can be indicative of high-temperature heating. The relatively low oxygen content of organic matter of UCAMMs is also consistent with a limited heating during atmospheric entry.

Regarding the mineral content of UCAMMs, high temperature heating effects are less constrained than by organic matter, when considered apart from the fact that the mineral assemblages in UCAMMs are embedded in the organic matter. Bradley et al. (2014) experimented the effect of heating on IDPs and MMs. Heating MMs around 500°C leads to the mobilization of sulfides within amorphous silicate that will accumulate on the grain surface. Interaction with atmospheric $O_2$ may oxidize it to magnetite rim if temperatures reach 1200°C (Toppani et al., 2000). Heating experiments also show the formation of large Fe-sulfides (hundreds of nm) decorating silicate, although they were thought to result from gas-phase sulfidization in the solar nebula (Keller and Messenger, 2011). In UCAMMs, although primary minerals often display large decorative Fe-sulfides similar to those formed in heating experiments (Bradley and Ishii, 2017), no clear magnetite rim has been observed. Some peripheral iron oxides, with stoichiometry close to magnetite are observed in several mineral assemblages. The observed Fe-oxides could result from mild heating during atmospheric entry processing, also indicated by a possible moderate thermal processing of fine-grained inorganic inclusions. The crystallization of chondrule-like material (see Dobrică et al. 2012, and Figure 12) can however not have occurred during atmospheric entry flash-heating, which would have destroyed the organic content around it. In DC06-308, Na is below detection limit, whereas the Na/Si ratio is about chondritic in DC16-309. The Na depletion in DC06-308 can thus be explained by mild atmospheric entry heating. The presence of a potential phyllosilicate in DC16-309 also sets an upper limit to the reached temperature (<600°C) during atmospheric entry, as phyllosilicates are rapidly thermally decomposed (Greshake et al., 1998). The ubiquitous presence of Fe sulfides within UCAMMs mineralogy indicates an upper limit around 650°C for atmospheric entry heating temperature as they are easily removed during high temperature oxidizing event.

## 4.4 On the possibility of aqueous alteration of cometary particles

### 4.4.1 Aqueous alteration of minerals

The secondary mineral phases such as a phyllosilicate candidate (this work) and carbonates (Dobrică et al., 2012) in UCAMMs could have formed out of hydrothermal alteration on a first parent body in the inner regions of the protoplanetary disk before being incorporated into comets. Whether secondary minerals observed in cometary materials are in-situ alteration products or heritage from asteroidal hydrothermal alteration is questioned here.

Internal heating and radiogenic decay are the main sources of energy to explain liquid water formation and alteration on asteroid surfaces. However, if the cometary parent body internal heating or impact-generated heating are not efficient enough, an alternative scenario has to be proposed to explain the potential water activity on comets. Suttle et al. (2020) proposed that aqueous alteration mostly occurs when silicates interact directly with liquid water during the comet's perihelion passage, generating enough heat and/or pressure to maintain the presence of liquid water on the subsurface of the cometary body. This process could explain the rare presence or detection of hydrated mineral in cometary dust. However not all comets meet the conditions to sustain liquid water for long enough timescale during perihelion passage. According to Suttle et al. (2020), only about 6% to 9% of periodic comets could host the necessary conditions to trigger enough subsurface aqueous alteration.

Experimental aqueous alteration of CP-IDPs showed that aqueous alteration of silicate could occur in interaction with either liquid or water vapor (Nakamura-Messenger et al., 2011; Takigawa et al., 2019). When directly exposed to liquid water at basic pH and relatively low-temperatures (25°C to 160°C), amorphous silicates like GEMS can be altered on timescale of hours. The alteration of crystalline silicates like enstatite could require longer time scales or higher temperatures. Extrapolation to lower temperatures kinetics indicates that alteration could occur at around 0°C over extended timescales (on the order of weeks). The reaction rate when silicates are exposed to water vapor phases is much slower (100 -120 hours) as it is only driven by the formation on a thin water film at the surface of silicates if pressure and temperature are high enough. When comparing reaction rates of liquid/solid interaction and vapor-solid interaction, it is unlikely that water vapor led to strong alteration process in cometary dust although its orbital window is wider than for liquid water (Suttle et al., 2020).

Concordia micrometeorites are collected by melting snow and filtering water (Duprat et al. 2007). It is necessary to examine a potential aqueous alteration during the collection of the UCAMMs in Dome C. During the extraction process, particles can stay in Antarctica water for up to 15 hours, at temperature lower than 20°C. Antarctic snow and water have variable pH comprised between 4.8 and 6.1 (Ali et al., 2010; Fujii, 1983). Nakamura-Messenger' experiments (2011) showed that pH is critical to the kinetic of alteration reaction. Their experiments were carried at higher pH values, to simulate cometary surface conditions (pH=11). Such large difference in acidity would critically slow down the reaction kinetic in the case of alteration with Antarctic water. Moreover, if aqueous alteration was dominant in Antarctica during the process of the Concordia collection, all samples would be extensively hydrated, which is not the case.

Yabuta et al. (2017) have proposed a formation of UCAMMs' organic matter through slight aqueous alteration. Mineralogical characteristics of UCAMMs tend to show minor aqueous alteration effects such as the low metal concentration in GEMS-like objects found in UCAMMs (Dobrică et al., 2012). Even with limited aqueous alteration, metals phases are the most easily mobilized phases and can be quickly depleted. The presence of Fe sulfides with compositions approaching that of pentlandite (Figure 14B) points toward a small amount of hydrothermal alteration. Their localization at the periphery of silicate minerals is compatible with a small amount of fluid circulation within the carbonaceous matrix of UCAMMs. The presence of a phyllosilicate-like mineral exhibiting a fibrous texture in UCAMM DC16-309 is also attesting of an aqueous alteration event (Figure 13A). The phyllosilicate candidate is enriched in Al and decorated with large Fe sulfides. Its composition is close to that of aqueously altered GEMS at basic pH obtained by Nakamura-Messenger et al (2011). The formation of DC16-309 phyllosilicate candidate could therefore result from an aqueous alteration in conditions close to those sustainable on cometary surface (pH=12, T=25°C) during perihelion. The heritage from a previously altered body can however not be discarded.

Characterization of organic and inorganic features from increasingly altered micrometeorites by Noguchi et al. (2017) shows that the global mineralogy and organic content of UCAMM is compatible with the those observed in slightly aqueously altered AMMs. These similarities suggest that UCAMMs' mineral content could have possibly evolved through a weak aqueous alteration process on their parent body.

### 4.4.2 Organic matter aqueous alteration vs heterogeneous organic precursors

If UCAMMs experienced aqueous alteration, their organic content should display chemical evidence of this process. The effect of hydrothermal alteration on insoluble organic matter (IOM) of carbonaceous chondrite as a part of its formation mechanism has already been described in several works (Alexander et al., 2007; Le Guillou et al., 2014; Vinogradoff et al., 2017). Progressive hydrothermal alteration leads to increased aromatization of the organic carbon associated with the loss of aliphatic carbon, although the H/C ratio and the alteration degree of IOM do not show correlation for several CMs chondrite (Alexander et al., 2007; Alexander et al., 2010). Alteration of IOM also leads to an increase in the concentration of oxygen within the organic content through chemical oxidation. It might lead to decrease in N/C and H/C ratios of the organic content as it gets oxidized, which is not necessarily observed in carbonaceous chondrite IOM of increasing oxidized types.

Among the three different types of OM that have been identified in UCAMMs (types I, II and III), type I and II share clear characteristics with IOM of CCs while type III organics of UCAMMs seems different. It has already been proposed that in UCAMMs type I and II on one hand, and type III OM on the other hand, could have formed through different processes. Type I and II, associated with crystalline phases, may have formed in the same way as IOM, in region close to the sun and could have then experimented hydrothermal alteration on a first parent body. A process is needed to explain the spatial configuration of type I and II OM in UCAMMs' organic phases. Orthous-Daunay et al. {, 2013 #4162} investigated the sources of variability in chondritic IOM and concluded that two processes could account for the observed variations. Either the organic precursors originally accreted were chemically distinct leading to the observed heterogeneity or the organic matter was modified by short-duration thermal processing on an asteroid-like parent body. These two processes are not mutually exclusives and some objects could have experimented both. In the case of UCAMMs, the observed configuration for type I and II most likely results from the accretion of different precursors as a formation only through a short-duration thermal event could hardly generate such a reproducible pattern in different samples. However, the possibility of a subsequent aqueous alteration episode of the heterogeneous organic matter cannot be ruled out and could have further emphasized the chemical distinction between type I and II OM with the potential redistribution of organics through fluid-interactions {Le Guillou, 2014 #3144}. The formation scenario for type III OM is different. It is suggested to have formed out of irradiation of N-rich ices in distant solar system region (Augé et al., 2015). Such formation in the outer region of the solar system suggest that type III OM in UCAMMs formed after the IOM-like organics. This suggest efficient transport mechanism in the young protoplanetary disk, where the parent body that originally accreted type I, type II and high temperature phases was brought to the outer region the solar system, beyond the nitrogen snow line. Whether these three phases experimented a subsequent aqueous alteration during comet's perihelion is still uncertain as several alteration steps could overlap in their signature. Alteration on a primitive parent body (as it could be for type I and II OM) and alteration on the cometary surface during perihelion occurs under very different temperature and pressure conditions and the effect of a second alteration in cometary surface conditions is thought to have a limited effect on the chemical signature of organics (Orthous-Daunay et al., 2013).

## 4.5 On the formation of UCAMMs

UCAMMs are assembling components formed at different locations in the protoplanetary disk and their chronological history of formation might be complex to unveil. The crystalline component of UCAMMs such as olivines and pyroxenes formed at high temperature, likely in the inner solar system. The presence of chondrule-like objects in UCAMMs also points out to a formation of crystalline minerals (or mineral assemblages) in the inner regions of the Solar System. The crystalline minerals could not be inherited directly from the ISM as interstellar dust is largely amorphous (Kemper et al., 2004, 2005; Wright et al., 2016). These crystalline phases are associated and embedded within organic matter of type I and II in UCAMMs, indicating a potential close-by formation location and/or timing, after the high temperature crystalline phases. These mixtures of organics and mineral could have experimented fluid-interactions during the first steps of their formation and water circulation on the surface of a primitive parent body could explain some of the alteration features seen in some UCAMMs. This material then needs to be transported toward the outer solar system to explain the presence of minerals fragment (and of their associated organic matter) in cometary dust. In this cometary forming region, UCAMMs parent bodies could have accreted N-rich ices on theirs surfaces. Irradiation of these ices with energetic particles such as Galactic Cosmic Rays can lead to the formation of an organic residue that would form large patches of N-rich type III OM.

From there, UCAMMs parent bodies could have started their cometary journey and at some moment being perturbed gravitationally and travel towards in the inner solar system. During their perihelion, conditions on their surface could have been sufficient to maintain liquid water so that aqueous alteration might have occurred. However, it is hard to differentiate the original alteration (during off just after the first step of mineral formation) from the late ones, in this scenario, during the travel toward the inner solar system. If cometary particles exhibit components indicating a potential aqueous alteration, it is not clear whether it is a heritage from early hydrothermal alteration on a asteroidal parent body or a product from aqueous alteration event during comet's perihelion.

# 5 Summary

Combined STXM-XANES and STEM analyses provide information on the organic and inorganic content of UCAMMs. The 8 UCAMMs presented here are mostly made of organic matter, extending over several tens of $\mu m^2$, with minor embedded mineral phases.

STXM-XANES analysis of the organic matter shows the presence of 3 distinct organic phases in UCAMMs. Type I and II organic matter (OM) have spectra resembling that of meteoritic IOM. The type III OM is rich in N (with N/C atomic ratios up to 0.2) and does not have counterparts in meteorites. Type I and II OMs can contain crystalline minerals and glassy phases, whereas type III OM contains glassy (GEMS-like) phases.

(S)TEM characterization of the mineral fraction of UCAMMs shows a complex and mostly unequilibrated mineralogy, with Mg-rich crystalline olivine and pyroxenes, Fe sulfides, Si-rich glassy phases, GEMS-like inclusions and minor mineral phases. A few hypocrystalline objects are also observed in UCAMMs. These mineral components are present either as extended aggregates as seen in DC06-18 or as isolated assemblages such as in DC06-308 and DC16-309. Overall, the individual assemblages have a Si-rich bulk composition with respect to CI, which is reminiscent of data from dust particles of comet 67P/Churyumov-Gerasimenko analyzed by COSIMA/Rosetta. STEM analysis revealed the presence of one mineral resembling a phyllosilicate in one UCAMM,

DC16-309. Whether this phase formed from aqueous alteration during comet's activity period, or if it was inherited from an already processed parent body is still unclear. A formation from alteration during the collection process in Antarctica is considered as unlikely. A small Na-rich phase with a composition close to $Na_2S$ was observed embedded in type II OM. GEMS-like phases seem rich in Na, and could represent a Na-rich reservoir of cometary matter.

Atmospheric entry heating did not seem to alter the UCAMMs in a harsh manner. There are evidences for a mild thermal event around 300°C that processed both the glassy phases and the organic matter.

At least a part of UCAMMs components were formed in the outer regions of the protoplanetary disk, as shown by their high abundance of OM, their high bulk N content and their high D/H ratios. However, crystalline minerals present in UCAMMs formed at high temperatures, close to the early Sun. Those high temperature phases could have experimented interaction with aqueous fluids on an asteroidal parent body They were later radially transported to the outer regions of the protoplanetary disk and incorporated UCAMMs definitive parent body. Whether this transport occurred in the form of a 'large' asteroidal parent body or under the form of smaller particles is still unclear. UCAMMs contain both unequilibrated crystalline minerals, Si-rich glassy phases and GEMS-like inclusion, and altered mineral phases such as a phyllosilicate-like mineral. The co-existence of pristine and altered inorganic phases within individual UCAMMs suggest accretion and mixing of these mineral phases that had diverse prior histories, rather than accretion of primitive phases and alteration on the UCAMM parent body.

The formation of OM probably happened in (at least) two steps, in order to explain the three kinds of OMs present in UCAMMs. The type I and II OMs share similarities with meteoritic IOM and might have the same origin. The type III OM could be formed by high-energy irradiation of N-rich ices, in the outer regions of the protoplanetary disk. The presence of a high nitrogen OM (with N/C atomic ratio ~0.5) inside GEMS-like inclusions would indicate that these glassy phases were incorporated in the icy UCAMM precursor before transformation of the N-rich ices into organic matter. The intimate mixing of the three kinds of OMs would either suggest a prior formation of types I and II OMs and transportation to the possibly N-rich ices that evolved upon irradiation into the type III OM, or co-formation of the three types of OMs from ices of different compositions. We favor the first hypothesis, as the type III OM only contains glassy inorganic phases that could have been accreted in situ from the presolar cloud. If type I and II OM formed from icy precursors together with type III OM, one would not expect the same mineral characteristics in the three types of OMs. The difference in area covered by the different OMs also points toward a two-step scenario. Type III OM is distributed over large area in UCAMM whereas types I and II, which are usually associated with each other, extend over smaller regions.

The observation of a Na-rich mineral phase in type II OM, and the presence of correlation between Na and N in the OM close to GEMS-like phases in one UCAMM could give insights into the host phase of Na in comets, questioning in particular the relevance of using $Na_2S$ in combination with ammonium salts to explain the reflectance spectra of comet 67P/Churyumov-Gerasimenko.

**Acknowledgments**: The micrometeorite collection was performed thanks to the logistic support of IPEV and PNRA. This work was supported in France by ANR COMETOR (ANR -18-CE31-0011-0), CNRS, LabEx P2IO, DIM-ACAV+ and CNES. We are also grateful to Bradley de Gregorio for providing Stardust STXM-XANES spectra.